\documentclass[fleqn,usenatbib]{mnras}

\usepackage[T1]{fontenc}
\usepackage{ae,aecompl}

\usepackage{graphicx}	
\usepackage{amsmath}	
\usepackage{amssymb}	
\usepackage{multirow}
\usepackage{comment}
\usepackage{pdflscape}
\usepackage{afterpage}

\newcommand{\Eq}[1]{Eq.~\ref{#1}}
\newcommand{\Fig}[1]{Fig.~\ref{#1}}
\newcommand{\Sec}[1]{Section~\ref{#1}}
\newcommand{\App}[1]{Appendix~\ref{#1}}

\newcommand{\threehun}{{\sc The Three Hundred}}

\DeclareRobustCommand{\VAN}[2]{#2}

\title[Infalling galaxy groups]{The Three Hundred project: Galaxy groups do not survive cluster infall}
\author[R. Haggar et al.]{\parbox{\textwidth}{
Roan Haggar,$^{1,2}$\thanks{E-mail: rhaggar@uwaterloo.ca}
Ulrike Kuchner,$^{1}$
Meghan E. Gray,$^{1}$
Frazer R. Pearce,$^{1}$}
\newauthor{Alexander Knebe,$^{3,4,5}$
Gustavo Yepes,$^{3,4}$ 
Weiguang Cui$^{6}$}
\\\\
\parbox{\textwidth}{
$^{1}$School and Physics \& Astronomy, University of Nottingham, Nottingham NG7 2RD, UK\\
$^{2}$Waterloo Centre for Astrophysics, University of Waterloo, Waterloo, Ontario N2L 3G1, Canada\\
$^{3}$Departamento de F\'isica Te\'{o}rica, M\'{o}dulo 15 Universidad Aut\'{o}noma de Madrid, 28049 Madrid, Spain\\
$^{4}$Centro de Investigaci\'{o}n Avanzada en F\'{\i}sica Fundamental (CIAFF), Universidad Aut\'{o}noma de Madrid, 28049 Madrid, Spain\\
$^{5}$International Centre for Radio Astronomy Research, The University of Western Australia, 35 Stirling Highway, Crawley, Western Australia 6009, Australia\\
$^{6}$Institute for Astronomy, University of Edinburgh, Royal Observatory, Edinburgh EH9 3HJ, UK
}}

\date{Accepted 2022 September 26. Received 2022 September 26; in original form 2022 March 29}

\pubyear{2022}

\begin{document}
\label{firstpage}
\pagerange{\pageref{firstpage}--\pageref{lastpage}}
\maketitle

\begin{abstract}
Galaxy clusters grow by accreting galaxies as individual objects, or as members of a galaxy group. These groups can strongly impact galaxy evolution, stripping the gas from galaxies, and enhancing the rate of galaxy mergers. However, it is not clear how the dynamics and structure of groups are affected when they interact with a large cluster, or whether all group members necessarily experience the same evolutionary processes. Using data from \threehun\ project, a suite of 324 hydrodynamical resimulations of large galaxy clusters, we study the properties of 1340 groups passing through a cluster. We find that half of group galaxies become gravitationally unbound from the group by the first pericentre, typically just $0.5-1$ Gyr after cluster entry. Most groups quickly mix with the cluster satellite population; only $8\%$ of infalling group haloes later leave the cluster, although for nearly half of these, all of their galaxies have become unbound, tidally disrupted or merged into the central by this stage. The position of galaxies in group-centric phase space is also important -- only galaxies near the centre of a group \mbox{($r\lesssim0.7R_{200}$)} remain bound once a group is inside a cluster, and slow-moving galaxies in the group centre are likely to be tidally disrupted, or merge with another galaxy. This work will help future observational studies to constrain the environmental histories of group galaxies. For instance, groups observed inside or nearby to clusters have likely approached very recently, meaning that their galaxies will not have experienced a cluster environment before.

\end{abstract}

\begin{keywords}
galaxies: clusters: general -- galaxies: groups: general -- galaxies: general -- methods: numerical
\end{keywords}



\section{Introduction}
\label{sec:intro}

Galaxy clusters grow by the gravitational accretion of smaller cosmic structures. These accreted structures vary in size, from individual galaxies, to galaxy groups containing tens or hundreds of member galaxies, to major cluster-cluster mergers involving thousands of galaxies \citep{moore1999,frenk2012}. Such hierarchical structure formation is one of the cornerstones of the Lambda cold dark matter ($\Lambda$CDM) model of the Universe \citep{white1978,navarro1996}. The wide range in sizes of their dark matter haloes, plus the existence of other structures such as cosmological filaments and walls \citep{bond1996,hahn2007}, results in a variety of cosmic environments in which galaxies can be found.

It is now well-established that the properties of galaxies strongly depend on where they are located. An early study by \citet{dressler1980} revealed that cluster environments contain mostly early-type galaxies, whereas galaxies in field regions typically have late-type morphologies\footnote{This idea had been noted previously in other works, such as \citet{hubble1936} and \citet{zwicky1937}.}. Furthermore, cluster galaxies have quenched star formation rates \citep{balogh1999,mcnab2021} and lower gas fractions \citep{jaffe2015} compared to field galaxies, across a large range of redshifts \citep{quadri2012}. Numerous mechanisms can explain this difference in gas content and star formation rate, including slow quenching processes such as galaxy starvation \citep{larson1980,maier2016,maier2019}, and rapid processes such as ram pressure stripping \citep{gunn1972,abadi1999,zabel2019}.

Although they represent the densest, most extreme galaxy environments, clusters are not the only structures that can dramatically impact galaxy evolution. Intermediate density environments can also play an important role: for instance, galaxy groups have been shown to enhance the rate of galaxy mergers, due to their combination of a high galaxy number density, and low velocity dispersion\footnote{The relative velocities of merging galaxies are usually \mbox{$<500$ km s$^{-1}$} \citep{lotz2008,an2019}, but dark matter haloes with masses greater than \mbox{$10^{14}\ M_{\odot}$} typically have velocity dispersions greater than \mbox{$500$ km s$^{-1}$} \citep{mcclintock2019,wetzell2021}. Consequently, mergers are more likely in group-sized haloes, with masses less than \mbox{$10^{14}\ M_{\odot}$}.} \citep{jian2012}. Mergers drastically impact the evolution of galaxies, altering their morphology and potentially triggering outflows and AGN feedback that can remove gas.

A consequence of this connection between galaxies and their environments is that a galaxy's evolution is not just impacted by the environment in which it is currently found -- it can also be affected by the environments through which a galaxy has previously passed. In the context of clusters, `pre-processing' describes the environmental mechanisms that act on a galaxy before it is accreted by a cluster. For example, galaxies can enter a cluster through cosmological filaments, which can quench star formation similarly to clusters, albeit to a lesser degree \citep{kraljic2018,laigle2018}. This results in degeneracy, as it is not immediately clear whether cluster galaxies have been quenched by the cluster itself, or are quenched due to pre-processing. However, it is clear that these filaments are an important factor to consider: for instance, \citet{kuchner2022} found that $45\%$ of cluster galaxies are accreted via filaments.

As galaxies can also enter clusters as members of galaxy groups, these are another contributor to pre-processing, although the exact degree of groups' contribution is debated. Some simulations \citep{mcgee2009,han2018} and observations \citep{dressler2013} find that close to half of all cluster members have been accreted as members of galaxy groups, while others \citep{arthur2019} find a much lower fraction. There are multiple explanations for this. For example, previous studies have shown that this fraction depends on the stellar mass of the accreted galaxies \citep{delucia2012}, and whether hydrodynamical or $N$-body simulations are used \citep{haggar2021}. Additionally, the definition of a `galaxy group' is not standardised, and different definitions can lead to different conclusions. Various studies have identified group members as galaxies that lie within the radius of a host group halo \citep{arthur2019,donnari2021}, that satisfy a boundness criterion \citep{han2018,choquechallapa2019}, or by using a Friends-of-Friends algorithm \citep{benavides2020}, all of which can result in different selections of group members. Furthermore, \citet{berrier2009} found that, although $30\%$ of cluster members (with dark matter halo masses greater than \mbox{$10^{11.5}\ h^{-1}M_{\odot}$)} are accreted via group haloes, half of these `groups' only contain two or three galaxies. Clearly, the minimum (and maximum) size of what constitutes a group is also an important consideration.

Both theoretical and observational studies have shown that the effects of a group environment on the evolutionary processes in galaxies can be enhanced even further when a group enters a cluster. Galaxy mergers \citep{benavides2020} and gas removal \citep{pallero2019,kleiner2021} are common in infalling groups, and multiple studies have connected this galaxy evolution to the external forces acting on a group, such as the effects of large-scale structure and clusters. \citet{vijayaraghavan2013} used cosmological simulations to show that mergers, ram pressure stripping, and tidal truncation of galaxy haloes are all enhanced further when their groups enter clusters, for a variety of reasons -- for example, the intra-group medium is shocked during a group-cluster merger, increasing its density and thus increasing the ram pressure stripping of the group members. Similar mechanisms have been described in previous works, such as \citet{mauduit2007}, who showed that near the centres of clusters lying in the core of the Shapley Supercluster, galaxies have lower radio loudness than galaxies elsewhere. They attributed this to the enhanced ram pressure stripping experienced by galaxies in shocked regions of merging clusters. In a related observational study, \citet{roberts2017} found that dynamically relaxed groups, which are typically isolated and slowly-growing, contain a smaller fraction of star-forming galaxies than unrelaxed groups. Again, this indicates that the processing of galaxies in groups is dependent on the disturbance of these groups by the larger environment in which they are located \citep[see also][]{gouin2021}.

All of this means that galaxies that have joined clusters as members of a group have experienced different evolutionary processes to those that have joined as individuals. \citet{bahe2019} used the Hydrangea suite of hydrodynamical simulations \citep{bahe2017} to study the survival fractions of galaxies entering clusters -- in their case, galaxies that do not `survive' are no longer resolved in the simulations, meaning they have either merged into a more massive galaxy (often a group central), or have been stripped below the total mass limit of \mbox{$5\times10^{8}\ M_{\odot}$}. \citet{bahe2019} showed that, after an infalling group enters a cluster, only $\sim50\%$ of its member galaxies survive to $z=0$. In contrast, they found that more than $90\%$ of galaxies that have not experienced any pre-processing survive to $z=0$. This survival fraction is higher than in some other studies, although much of the prior work in this field has used $N$-body simulations \citep[e.g.][]{gill2004_survival}, in which substructure can be more easily stripped \citep{smith2016}. The results of \citet{bahe2019} show that group members are particularly strongly influenced within clusters, and that they can be very heavily disturbed during accretion onto a cluster. 

Moreover, previous work has hinted that galaxy groups themselves can be heavily disrupted when entering a cluster. \citet{choquechallapa2019} found that, using dark matter-only simulations and a similar group definition as is used in this work, over $90\%$ of group members become unbound after a group enters a cluster, and that these galaxies quickly form part of the cluster population of galaxies. Furthermore, \citet{gonzalezcasado1994} showed that tidal forces from clusters can rapidly increase the internal energy of infalling groups, by up to a factor of 10 for the smallest groups. This can allow these groups to be disrupted, although it should be noted that absorbing more energy than the binding energy does not necessarily lead to the complete disruption of groups \citep{vandenbosch2018}.

However, beyond this, there is little work that has examined in detail how the dynamics of galaxy groups evolve when they are accreted by a cluster, particularly with large numbers of clusters in hydrodynamical simulations. While previous studies have looked at the overall disruption of groups that enter a cluster and the subsequent `post-processing' of their constituent galaxies, we do not currently have a detailed understanding of the timescales over which groups change, and how the evolutionary processes that galaxies experience are affected by the group dynamics \citep{cohn2012,bahe2019}.

In this work, we use \threehun\ project, a mass-complete sample of 324 galaxy clusters taken from a \mbox{$1\ h^{-1}$ Gpc} cosmological volume. These are resimulated out to distances of several times the $R_{200}$ of the cluster, where $R_{200}$ is the radius within which the mean density of a cluster is equal to 200 times the critical density of the Universe. We use these simulations to study the evolution of groups as they enter galaxy clusters, and the processes that galaxies in these groups experience in their subsequent passage through the cluster halo. Specifically, we look at how the phase space of groups evolves: that is, how the positions and speeds of galaxies change, relative to the group that they are bound to. We make comparisons between groups before and after they pass through a cluster, to find the cumulative effect that a cluster has on the dynamics and structure of galaxy groups. Then, we look at the fates of group galaxies, categorising them based on the processes they experience in the several Gyr after entering a cluster (such as mergers and stripping), and how this depends on the structure of groups. Finally, we discuss how this theoretical work can help observational studies.

The paper is structured as follows: In \Sec{sec:methods} we introduce the simulation data that we use, and the methods we use to analyse groups. In \Sec{sec:evolution} we show how the internal dynamics of groups change as they pass through a cluster, and in \Sec{sec:before_after} we focus on the state of galaxies and groups after passing through a cluster. Finally, we summarise our findings in \Sec{sec:conclusions}.

\section{Simulations \& Numerical methods}
\label{sec:methods}

Below we detail the methods and data used in this work. Much of this, particularly \Sec{sec:simulations} and \Sec{sec:subsample}, build on the analysis in our previous work, \citet{haggar2021}, in which we compare the substructure of galaxy groups and galaxy clusters in hydrodynamical and dark matter-only simulations.

\subsection{Simulation data}
\label{sec:simulations}

This work utilises data from \threehun\ project, a suite of 324 hydrodynamical resimulations of large galaxy clusters. The simulations were produced by extracting the 324 most massive clusters from the dark matter-only MDPL2 MultiDark simulation \citep{klypin2016}\footnote{The MultiDark simulations are publicly available from the cosmosim database, \url{https://www.cosmosim.org}.}, and resimulating each from its initial conditions with baryonic physics. This was done by taking all dark matter particles within \mbox{$15\ h^{-1}$ Mpc} of the cluster centre at \mbox{$z=0$} (between \mbox{$7-10R_{200}$} for the range of cluster masses in the sample), tracing the particles back to their initial positions, and then splitting each one into a dark matter and a gas particle, with masses set by the baryonic matter fraction of the Universe. Lower-resolution particles were used beyond \mbox{$15\ h^{-1}$ Mpc} to model any tidal effects of the surrounding large-scale structure. 

The MDPL2 simulation involves a box with sides of comoving length \mbox{$1\ h^{-1}$ Gpc}, simulated using \textit{Planck} cosmology (\mbox{$\Omega_{\rm{M}}=0.307$}, \mbox{$\Omega_{\rm{B}}=0.048$}, \mbox{$\Omega_{\Lambda}=0.693$}, \mbox{$h=0.678$}, \mbox{$\sigma_{8}=0.823$}, \mbox{$n_{\rm{s}}=0.96$}) \citep{planck2016}. The same box size and cosmology are used for each of the cluster simulations in \threehun, so that each cluster is embedded in a comoving box of size \mbox{$1\ h^{-1}$ Gpc}, most of which is occupied by the low-resolution particles described in the previous paragraph. Consequently, the lengths and distances quoted throughout this work are also given in comoving coordinates.

The hydrodynamical resimulations were carried out using the {\sc GadgetX} code. {\sc{GadgetX}} is a modified version of the {\sc{Gadget3}} code, which is itself an updated version of the {\sc{Gadget2}} code, and uses a smoothed-particle hydrodynamics scheme to fully evolve the gas component of the simulations \citep{springel2005_gadget2, beck2016}. The final dataset comprises of a mass-complete cluster sample from \mbox{$M_{200}=5\times10^{14}\ h^{-1}M_{\odot}$} to \mbox{$M_{200}=2.6\times10^{15}\ h^{-1}M_{\odot}$}, where $M_{200}$ is the mass contained within a sphere of radius $R_{200}$. The dark matter and gas particles in the simulations have masses of \mbox{$m_{\rm{DM}}=1.27\times10^{9}\ h^{-1}M_{\odot}$} and \mbox{$m_{\rm{gas}}=2.36\times10^{8}\ h^{-1}M_{\odot}$} respectively. The simulations also contain stellar particles of variable masses, typically with \mbox{$m_{\rm{star}}\sim4\times10^{7}\ h^{-1}M_{\odot}$}, produced by the stochastic star-formation model that is implemented by {\sc{GadgetX}} \citep{tornatore2007,murante2010,rasia2015}. A Plummer equivalent gravitational softening length of \mbox{6.5 $h^{-1}$ kpc} is used for the dark matter and gas particles, and \mbox{5 $h^{-1}$ kpc} for the stellar particles. \threehun\ dataset is described in more extensive detail in \citet{cui2018}, and has been used in numerous previous studies to examine galaxy groups \citep{haggar2021}, environment \citep{wang2018}, cosmic filaments \citep{kuchner2020,rost2021,kotecha2022}, backsplash galaxies \citep{haggar2020} and ram pressure stripping \citep{arthur2019,mostoghiu2021}, among other areas. The full simulation suite also includes simulations with different physics models, however in this work we only use the {\sc{GadgetX}} simulations.

\subsubsection{Galaxy identification and tree-building}
\label{sec:tree}

The data for each cluster in \threehun\ consists of 129 snapshots saved between \mbox{$z=16.98$} and \mbox{$z=0$}, separated by approximately \mbox{$0.3$ Gyr} at low redshift. To identify the haloes and subhaloes, each snapshot was processed using the Amiga Halo Finder, {\sc ahf}\footnote{\url{http://popia.ft.uam.es/AHF}} (see \citet{gill2004_ahf} and \citet{knollmann2009} for further details). {\sc{ahf}} operates by identifying peaks in the matter density field, and returns the positions and velocities of haloes and subhaloes, as well as their radii, their mass in gas, stars and dark matter, and a host of other properties.

The halo merger trees were built using {\sc mergertree}, a tree-builder that forms part of the {\sc ahf} package. For each halo in a given snapshot, this tree-builder calculates a merit function with respect to all haloes in previous snapshots; specifically, {\sc mergertree} uses the merit function $M_{\rm{i}}$, as described in Table B1 of \citet{knebe2013}. This merit function is then used to identify a main progenitor, plus other progenitors, based on the number of particles that they share with the halo of interest. The tree-builder has the ability to skip snapshots, and thus is able to `patch' over gaps in the merger tree, for example when a subhalo is near to the centre of its host halo and so is not easy to identify against the high background density \citep{onions2012}. We also place a limit on the change in mass permitted between successive snapshots, such that no halo can more than double in dark matter mass. This helps to prevent `mismatches', caused by a subhalo located close to the centre of a larger halo being detected as the main halo \citep[as shown in][]{behroozi2015}. Additional information on {\sc ahf} and {\sc mergertree} can be found in \citet{knebe2011_ahf} and \citet{srisawat2013}.

\subsection{Sub-sample of clusters}
\label{sec:subsample}

Some of the clusters exhibit some minor problems in the trees constructed by {\sc mergertree}, which we describe below. However, thanks to the large dataset that we are using, we can identify and remove these objects, and still be left with a large sample of simulated clusters.

In some cases the merit function used by {\sc mergertree} can incorrectly assign links between haloes in different snapshots. This can lead to an apparent `jump' in the position of a halo or subhalo (in box coordinates), as well as a sudden change in its properties, due to one halo being incorrectly labelled as the progenitor of another. These mismatching events are uncommon, typically only affect a small number of snapshots, and are fairly inconsequential when they affect individual galaxy haloes. However, the merger tree of the main cluster halo can also be affected in this way, leading to a sudden change in the position of the main halo. Such a change in position is particularly problematic in this work, because it will result in many galaxies and groups being erroneously tagged as members of a cluster. 

These merger tree mismatches are especially common during a major merger between two haloes. \citet{behroozi2015} showed that various halo finders experience this same problem, where two merging haloes of similar size can be accidentally switched by a tree-builder, leading to the sizes and positions of haloes appearing to change suddenly and dramatically. Many of the clusters in \threehun\ experience major mergers; a recent study, \citet{contrerassantos2022}, discusses cluster mergers in \threehun\ simulations in detail. In fact, we find that 59 of our 324 simulated clusters experience a change in position of \mbox{$>0.5R_{200}(z)$} between two snapshots after \mbox{$z=1$}. We find that, given that the typical time elapsing between snapshots at this redshift is \mbox{$\sim0.3$ Gyr}, this distance is non-physical and so likely due to these tree-builder issues.

In some cases, the tree-builder instead misses a link in the merger tree, causing a branch of the merger tree to end prematurely and the history of the halo before this link to be lost. For 17 clusters, the central cluster halo is affected in this way, and the evolution of the cluster halo cannot be tracked back further than $z=0.5$. We choose to also remove these clusters from our analysis, in order to avoid affecting our results with clusters that do not have complete, reliable merger trees. Nine of these clusters also experience the halo mismatches described in the previous paragraph, resulting in a total of 67 clusters that we choose to remove from our sample.

The remaining 257 clusters have $M_{200}$ masses (dark matter, gas and stars, including subhaloes) ranging from \mbox{$5\times 10^{14}\ h^{-1}M_{\odot}$} to \mbox{$2.6\times 10^{15}\ h^{-1}M_{\odot}$}, with a median value of \mbox{$8\times 10^{14}\ h^{-1}M_{\odot}$}. Their radii ($R_{200}$) range from \mbox{$1.3\ h^{-1}$ Mpc} to \mbox{$2.3\ h^{-1}$ Mpc}, with a median of \mbox{$1.5\ h^{-1}$ Mpc}.

\subsection{Galaxy and group selection}
\label{sec:galgroups}

In this work, we place lower limits on the total mass (including dark matter, gas and stars) and the stellar mass of galaxies in the simulations, so that all the haloes we keep from our halo finder represent real, physical galaxies. We only examine galaxy haloes with a total mass of \mbox{$M_{200}\geq10^{10.5}\ h^{-1}M_{\odot}$}, which corresponds to approximately $100$ particles in the high-resolution regions containing the clusters. We also only use galaxies with a stellar mass $M_{\rm{star}}\geq10^{9.5}M_{\odot}$. We consider these to be physical galaxies that have built up a substantial population of stars -- this cut is approximately equivalent to removing all galaxies with a luminosity \mbox{$L<10^{8}L_{\odot}$}, whilst keeping all galaxies with \mbox{$L>10^{9}L_{\odot}$}. This stellar mass cut also allows us to investigate a similar population of galaxies to upcoming observational studies, such as the WEAVE\footnote{\url{https://www.ing.iac.es//confluence/display/WEAV}} Wide-Field Cluster Survey, which will study cluster galaxies down to stellar masses of \mbox{$\sim10^{9}\ M_{\odot}$} \citep[e.g.][]{kuchner2020}. Finally, we remove all galaxies from our simulations that contain more than $30\%$ of their mass in stars. These objects are generally found extremely close to the centre of a larger halo, and so have been heavily stripped \citep{knebe2020}, leaving remnants with high stellar mass fractions, whose properties (such as their radii and masses) are not well-defined by our halo finder. These objects are very rare, and make up only $1\%$ of all haloes within $5R_{200}$ of the clusters, so we make the decision to remove these objects from our analysis. By applying these three constraints to our simulations, we consider all remaining objects to be realistic galaxies with a significant population of stars at $z=0$.

\subsubsection{Group identification}
\label{sec:group_id}

Throughout this work, we identify galaxy groups by taking each galaxy, assuming its halo to be the host halo of a galaxy group, and then determining if any other galaxies in the same snapshot are associated with it. We identify galaxies as being associated with a halo (and thus members of the group) using the same approach as \citet{han2018}. They assume that a group's dark matter halo follows a spherically symmetric NFW density profile \citep{navarro1996}, truncated at $R_{200}$. Using this to calculate the gravitational potential of the group halo, they identify group members as those that satisfy the criterion given below:

\begin{equation}
    \frac{v^2}{2}+\Phi\left(r\right)<\Phi\left(2.5R_{200}^{\rm{grp}}\right)\,.
    \label{eq:bounded}
\end{equation}

Here, $v$ is the relative velocity of a galaxy with respect to its group host, $\Phi(r)$ is the gravitational potential due to the group host at a distance $r$ from its centre, and $R_{200}^{\rm{grp}}$ is the radius of the group host halo. It is important to note that this is different to the radius of the host cluster in each simulation, which is subsequently referred to by $R_{200}^{\rm{clus}}$. Any galaxies that are less massive than their group host and that satisfy \Eq{eq:bounded} are taken to be bound members of this group. Although we hereafter refer to these group members as being `bound' to their host group, it is important to note that this definition is not technically equivalent to gravitational binding. Previous work \citep[e.g.][]{behroozi2013} has shown that halo particles can be gravitationally balanced against the Hubble flow out to $\sim4R_{200}$ from the halo centre. However, \Eq{eq:bounded} places an artificial radial limit on groups, so that galaxies can only be found as far as $2.5R_{200}^{\rm{grp}}$ from the centre of the group.

This outer limit is the same as was used by \citet{han2018}: their choice was motivated by the work of \citet{mamon2004}, who showed that backsplash galaxies are typically found out to approximately $2.5R_{200}$ from their host halo, but rarely any further. By setting this as the outer limit of a group, we include almost all galaxies that are on bound orbits around the group (having passed through its central halo at least once), whilst excluding galaxies that have not entered the group halo before. Furthermore, the relative velocity term in \Eq{eq:bounded} means that only slow-moving galaxies at large distances are included as group members. Galaxies moving at greater velocities are excluded from the group, as these are likely `fly-by' galaxies or `renegade subhaloes' \citep{knebe2011_renegade}, which happen to be passing near to the group, but are not bound to it.

If a halo has four or more galaxies associated with it that each have a smaller total mass (including dark matter, gas, and stars) than the halo, we define this as a group, with the halo being the `group host' halo. Throughout this work we assume that the central group galaxy in each of these group host haloes exists at the centre of the halo, which has been shown to be the case in previous work. \citet{lin2004} used X-ray observations of groups and clusters with masses similar to those in this work (\mbox{$10^{13.5}\ M_{\odot}<$}\mbox{$\ M_{200}<\ $}\mbox{$10^{15.3}\ M_{\odot}$}) to show that in $75\%$ of these haloes, the brightest galaxy is located within $0.06R_{200}$ of the halo centre; this result is corroborated by both \citet{hwang2008} and \citet{stott2012}. 

Very small groups with fewer than five members are common, but for the mass constraints that we put in place in \Sec{sec:galgroups}, a collection of $\gtrsim5$ associated galaxies is typically required to define a group \citep[see][for example]{tully2015}. Additionally, we only study groups with 50 or fewer members, as detailed in the following section. Again, we stress that this limit applies to the number of group members that satisfy the mass constraints in \Sec{sec:galgroups}, as is the case throughout the rest of this work unless stated otherwise.

\subsubsection{Infalling groups}
\label{sec:infalling}

The focus of this work is on the evolution of galaxy groups as they enter and pass through a galaxy cluster. In order to study this, we identify a sample of infalling galaxy groups at all redshifts, in the same way as our previous study \citep{haggar2021}, and other previous work \citep[e.g.][]{choquechallapa2019}. To do this, we identify all galaxies that have just fallen into the cluster; these are galaxies that are within $R_{200}^{\rm{clus}}$ of the cluster centre, having been outside of the cluster in the previous snapshot. These objects are referred to as the `infalling' galaxies. Note that we do not distinguish by the time at which the galaxies entered the cluster -- these infall events can happen at any time over a cluster's history. 

We then examine each of these galaxies using the method described in \Sec{sec:group_id}, to determine whether each object is the host halo of a galaxy group that has passed within the radius of the cluster. We keep groups with between five and 50 members (including the host object) that each satisfy the mass constraints given in \Sec{sec:galgroups}. Groups of this richness are considered to be small or intermediate sized groups \citep{tully2015}, but are large enough to provide an environment that can strongly impact galaxy evolution \citep{hester2006}. Because of the upper limit of 50 members on the group size, major cluster-cluster mergers are not included in this study.

\begin{figure}
	\includegraphics[width=\columnwidth]{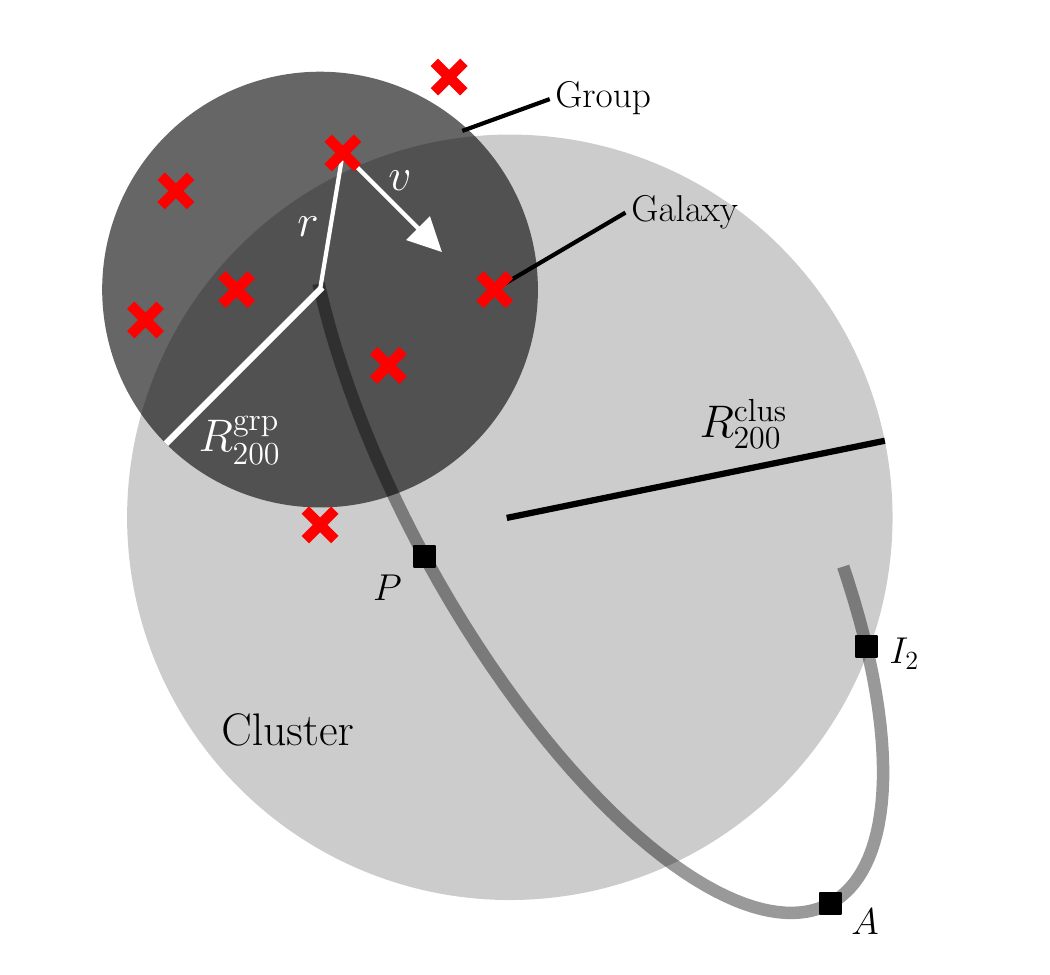}
    \caption{Schematic of a galaxy group halo (dark circle) passing within $R_{200}^{\rm{clus}}$ of a cluster (light circle) for the first time. Red crosses represent galaxies that are members of this group; note that these are not limited to be within $R_{200}$ of the cluster or the group at infall, but are just defined based on \Eq{eq:bounded}. The position, $r$, and velocity, $v$, of one galaxy relative to its host group are also labelled. The subsequent path of this group through the cluster is shown by the thick, grey line, and the black squares on this line represent the moments of pericentre, apocentre, and second infall of the group (marked $P$, $A$ and $I_{\rm{2}}$ respectively), which are used extensively in \Sec{sec:before_after} of this work.}
    \label{fig:schematic}
\end{figure}

\Fig{fig:schematic} shows a schematic view of a galaxy group at the moment of infall, and its subsequent passage through a cluster. Finally, we exclude any groups that have passed through the cluster once previously, so that all of the groups in our sample are entering a cluster for the first time. Groups on a second (or subsequent) infall make up less than $1\%$ of the groups we identify, so we assume that this will not strongly impact our results.

Overall, we identify 1340 infalling groups across the 257 clusters that we use in this work, with a median richness (number of galaxies) of \mbox{$8^{+7}_{-3}$} members ($1\sigma$ spread). This indicates that, although we permit groups to contain up to 50 members, groups of this richness are rare compared to the large number of poorer groups -- only $8\%$ of the groups contain more than 20 members. The average mass, $M_{200}^{\rm{grp}}$, of these groups at cluster infall is \mbox{$10^{13.5\pm0.4}\ h^{-1}M_{\odot}$} (median and $1\sigma$ spread). This means that the typical mass ratio between a group and cluster is roughly \mbox{$1:20$}, although this varies across the range of group and cluster masses, from approximately \mbox{$1:5$} to \mbox{$1:100$}. Finally, these groups enter the cluster over a wide range of redshifts, with a median value of \mbox{$z_{\rm{infall}}=0.4^{+0.6}_{-0.3}$}.

\subsubsection{Tidal radius of groups}
\label{sec:tidal_radius}

Subhaloes passing through a larger halo can experience strong tidal stripping, and group-sized haloes can often lose a large fraction of their mass due to stripping from a cluster \citep{muldrew2011, bahe2019}. Similarly, galaxies can be tidally stripped from these groups \citep{gonzalezcasado1994, choquechallapa2019}, although the extent of this stripping varies between different studies. For instance, \citet{vijayaraghavan2015} found that the central regions of galaxy groups are largely unaffected by a cluster potential, and are only disrupted by dynamical friction after several Gyr.

The tidal radius of a group or dark matter halo is an effective way to predict and explain tidal stripping. Generally, the tidal radius is defined as the distance from a smaller object at which the self-gravity of that object is less than the tidal force due to a larger object. However, the tidal radius is not precisely defined, and different definitions exist for different scenarios \citep[see][for a detailed summary]{vandenbosch2018}. Perhaps the simplest example is the Roche limit, the tidal radius of a point mass that is being tidally influenced by another point mass. More physically motivated scenarios such as an extended subhalo within a larger extended halo (as is used in this work) require more complex descriptions.

Calculating a tidal radius is complicated further by the fact that subhalo properties are often poorly defined by a subhalo finder, and can be strongly dependent on the distance of a subhalo from the group centre. \citet{muldrew2011} test the ability of {\sc {ahf}} and another halo finder, {\sc {subfind}} \citep{springel2001_subfind}, to recover subhalo properties. They find that {\sc {ahf}} performs better at identifying all the particles of a subhalo, and thus constrains the subhalo mass more effectively. However, for subhaloes within \mbox{$\sim0.5R_{\rm{vir}}$ ($\sim0.7R_{200}$)}\footnote{\citet{muldrew2011} use the definition of virial radius presented in \citet{bryan1998}. For their cosmology, the mean density of a halo within the virial radius is $101\rho_{\rm{crit}}$, where $\rho_{\rm{crit}}$ is the critical density of the Universe. Hence, for the clusters used in their work and ours, $R_{\rm{vir}}=R_{101}\approx1.3R_{200}$, although it is important to note that this conversion depends on the concentrations and density profiles of dark matter haloes.}, both halo finders underestimate the number of particles in the subhalo. This makes it challenging to predict the mass, and therefore the radius, of subhaloes in these regions. Furthermore, in our work we wish to combine the data from multiple galaxy groups (of different sizes) in multiple galaxy clusters (also of different sizes). It is therefore convenient to have an expression for the group tidal radius that is independent of the cluster or group size, and solely depends on the separation between these two.

We define the tidal radius of an infalling subhalo by adapting the descriptions in \citet{klypin1999} and \citet{vandenbosch2018}. Specifically, they give the tidal radius in terms of a function, $f(d)$, whose value is the minimum of two expressions:

\begin{equation}
    R_{\rm{t}}=d\left(\frac{M_{\rm{<d}}^{\rm{grp}}(R_{\rm{t}})}{M_{\rm{<d}}^{\rm{clus}}(d)}\frac{1}{2-f(d)}\right)^{\frac{1}{3}}\,,
    \label{eq:tidalrad1}
\end{equation}

\begin{equation}
    f(d)={\rm{min}}\left[1,\ \left.\frac{{\rm{d(ln}}M_{\rm{<d}}^{\rm{clus}})}{{\rm{d(ln}}d)}\right|_{d}\right]\,.
    \label{eq:tidalradchoose}
\end{equation}

Here, $R_{\rm{t}}$ is the tidal radius of the group, $d$ is distance from a group to the cluster centre, and $M$ are the radial enclosed mass profiles of the group and the cluster. We assume the radial density of the dark matter haloes follow an NFW profile \citep{navarro1996}, given by

\begin{equation}
    \rho(d)=\frac{\rho_{0}}{x\left(1+x\right)^2}\,,
    \label{eq:nfw}
\end{equation}

\begin{equation}
    x=\frac{d}{R_{\rm{s}}}\,,
    \label{eq:define_x}
\end{equation}
where $\rho(d)$ is the radial density of the halo in terms of the distance to its centre, $\rho_{0}$ is a characteristic density, and $R_{\rm{s}}$ is the scale radius of the halo. We also define the quantity $x$ to make the equations in this section more easily readable. The concentration of a halo, $c$, is equal to the ratio between $R_{200}$ and $R_{\rm{s}}$:

\begin{equation}
    R_{\rm{s}}=\frac{R_{200}}{c}\,.
    \label{eq:scaleradius}
\end{equation}

Integrating the NFW profile, \Eq{eq:nfw}, gives the enclosed mass in a sphere of radius $d$:

\begin{equation}
    M_{\rm{<d}}=4\pi\rho_{0}R_{\rm{s}}^{3}\left[{\rm{ln}}\left(1+\frac{d}{R_{\rm{s}}}\right)-\frac{d}{d+R_{\rm{s}}}\right]\,.
    \label{eq:nfw_enc}
\end{equation}

This can then be used to rewrite \Eq{eq:tidalradchoose}. For a general NFW profile, \mbox{$f(d)=1$} in the region \mbox{$d\lesssim2.2R_{\rm{s}}$}. However, \mbox{$f(d)<1$} outside of this region, and so must be calculated for each subhalo. Solving the derivative in the expression of $f(d)$ gives:

\begin{equation}
    f(d)={\rm{min}}\left[1,\ \left(\frac{(x/(1+x))^{2}}{{\rm{ln}}(x+1)-x/(1+x)}\right)\right]\,,
    \label{eq:tidalradchoose_approx}
\end{equation}
where $x$ is defined the same as in \Eq{eq:define_x}. 

Also using \Eq{eq:nfw_enc}, we can produce an expression for $M_{200}$, as \mbox{$M_{\rm{<d}}(d=R_{200})=M_{200}$}. Substituting this into \Eq{eq:nfw_enc} gives

\begin{equation}
    \begin{split}
    M_{\rm{<d}}=M_{200}&\left[{\rm{ln}}\left(1+x\right)-\frac{x}{1+x}\right]\\
    &\hspace{11pt}\times\left[{\rm{ln}}\left(1+c\right)-\frac{c}{1+c}\right]^{-1}\,.
    \end{split}
    \label{eq:nfw_enc2}
\end{equation}

This expression can then be substituted into the equation for tidal radius, \Eq{eq:tidalrad1}, for the cluster enclosed mass, $M_{\rm{<d}}^{\rm{clus}}(d)$, and for the mass enclosed within the tidal radius of a group, $M_{\rm{<d}}^{\rm{grp}}(R_{\rm{t}})$. This gives the expression for tidal radius below, 

\begin{equation}
    \begin{split}
    \frac{R_{\rm{t}}}{R_{200}^{\rm{grp}}}=\frac{d}{R_{200}^{\rm{clus}}}\left(\frac{1}{2-f(d)}\right)^{\frac{1}{3}}\ \ \ \ \ \ \ \ \ \ \ \ \ \ \ \ \ \ \ \ \ \ \ \ \ \ \ \ \ \ \\
    \times\left(\frac{\left[{\rm{ln}}\left(1+\frac{C_{\rm{c}}d}{R_{200}^{\rm{clus}}}\right)+\left(1+\frac{C_{\rm{c}}d}{R_{200}^{\rm{clus}}}\right)^{-1}-1\right]}{\left[{\rm{ln}}\left(1+C_{\rm{c}}\right)-\frac{C_{\rm{c}}}{1+C_{\rm{c}}}\right]}\right)^{-\frac{1}{3}}\\
    \times\left(\frac{\left[{\rm{ln}}\left(1+\frac{C_{\rm{g}}R_{\rm{t}}}{R_{200}^{\rm{grp}}}\right)+\left(1+\frac{C_{\rm{g}}R_{\rm{t}}}{R_{200}^{\rm{grp}}}\right)^{-1}-1\right]}{\left[{\rm{ln}}\left(1+C_{\rm{g}}\right)-\frac{C_{\rm{g}}}{1+C_{\rm{g}}}\right]}\right)^{\frac{1}{3}}\,,
    \end{split}
    \label{eq:tidalrad_final}
\end{equation}
where $C_{\rm{c}}$ and $C_{\rm{g}}$ are the concentrations of the cluster and group haloes, respectively, and $f(d)$ is given by \Eq{eq:tidalradchoose_approx}. Finally, we take the halo concentrations to be constant for all of the clusters, and all of the groups. Specifically, we set the value of $C_{\rm{c}}$ equal to the median value for our clusters, $C_{\rm{c}}=3.9$, and $C_{\rm{g}}$ equal to the median value for our groups, $C_{\rm{g}}=4.4$. Approximating these concentrations as constant has a small effect because \Eq{eq:tidalrad_final} is not strongly dependent on them. For a group at a distance $d=0.2R_{200}^{\rm{clus}}$ from the cluster centre, the value of $R_{\rm{t}}$ varies from its median value by $20\%$ across the full range of cluster concentrations (from $C_{\rm{c}}=2.3$ to $C_{\rm{c}}=7.7$). At greater distances from the cluster centre, this variation is even smaller. Similarly, the $1\sigma$ deviation\footnote{There are a small number of groups with highly concentrated haloes, so we use the $1\sigma$ spread in $C_{\rm{g}}$ to avoid skewing our data.} in $C_{\rm{g}}$, between $2.6$ and $6.9$, leads to a variation in $R_{\rm{t}}$ of less than $18\%$. This variation is used as the uncertainty in the tidal radii that we calculate in \Sec{sec:tidal_friction}.

By making these assumptions, we are able to reach an expression for the tidal radius of a group in units of $R_{200}^{\rm{grp}}$ that depends only on the distance from the group to the cluster centre. As $R_{200}^{\rm{grp}}$ of a group can change over the course of infall, the tidal radius could be scaled by this changing group radius. However, we instead choose to scale the tidal radius by $R_{200}^{\rm{grp}}$ at the moment of cluster infall, to allow us to stack groups and study their evolution more clearly -- this is explained in further detail in \Sec{sec:evol_infall}.

\Eq{eq:tidalrad_final} is an ideal form of the tidal radius for our analysis, as it allows us to calculate the average tidal radius for all groups in a radial bin across many clusters. This form of the tidal radius may also be useful in future studies, both observational and theoretical, that wish to stack substructure on multiple different size scales.

\section{Phase space evolution}
\label{sec:evolution}

Much of the work in this paper revolves around studying the phase space of galaxies within galaxy groups, as the groups enter and pass through a cluster, and how the distribution of galaxies within this phase space changes over time. This analysis follows the same basic process as in \citet{haggar2021}; the phase space consists of the radial distance of a galaxy from its host group halo in terms of the group halo radius, \mbox{$R_{200}^{\rm{grp}}$}, and the galaxy's velocity relative to the group halo, in units of \mbox{$v_{\rm{cir}}$}, the circular orbital velocity at \mbox{$r=R_{200}^{\rm{grp}}$}. It is important to stress that this work involves looking at the phase space of galaxies relative to their host group, not the cluster \citep[as has been done by numerous previous studies, such as][for example]{jaffe2015,arthur2019}. This method can provide detailed information, by showing both the spatial and velocity distribution of galaxies in groups, and telling us how the speed and acceleration of galaxies differ in different regions of the group.

\begin{figure}
\includegraphics[width=\columnwidth]{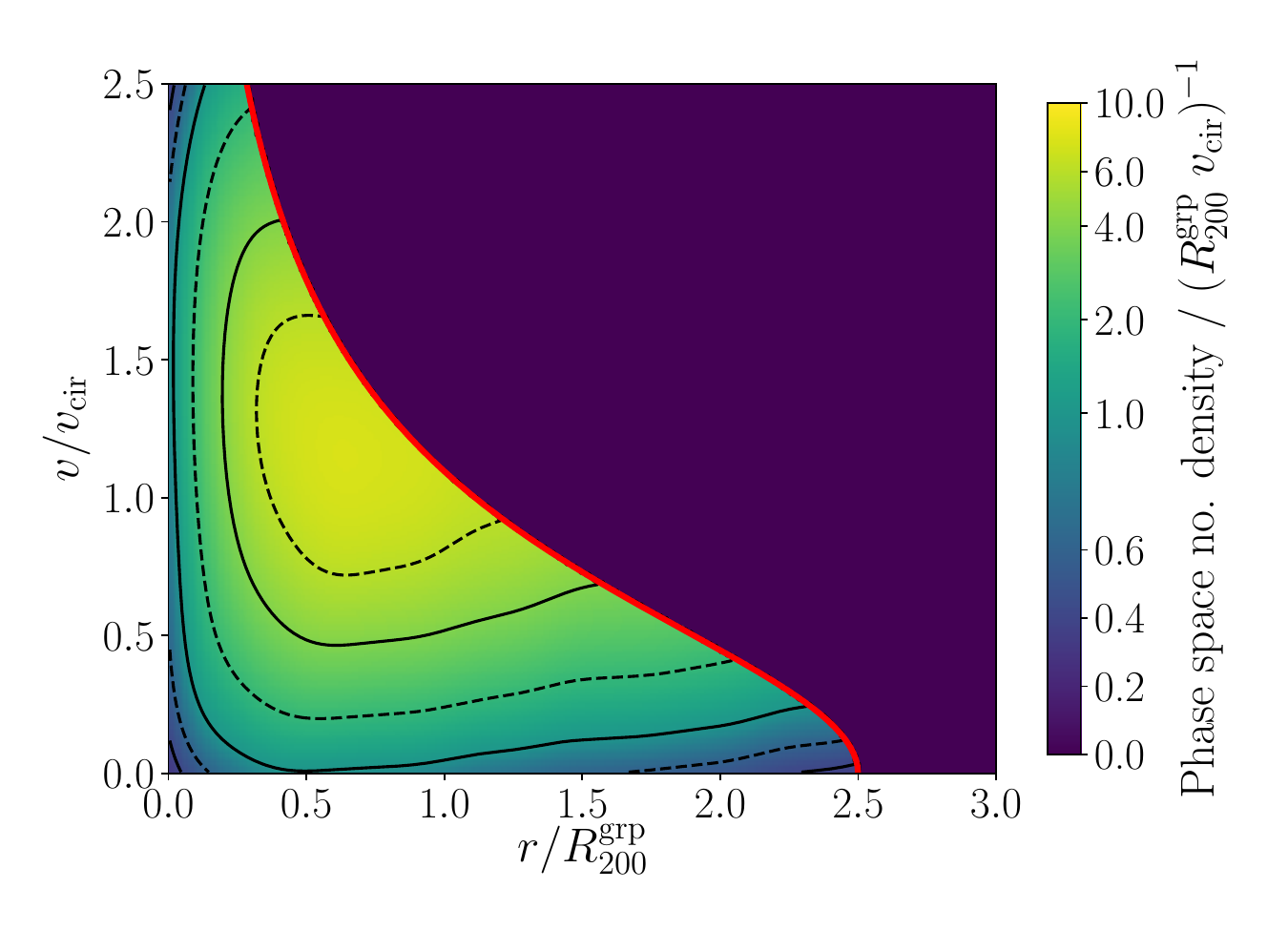}
\caption{Distribution of galaxies in group phase space, for groups at the moment of infall into the host cluster. Data for all groups from all 257 clusters that are used in this analysis are shown, stacked together. Lighter colours represent regions of phase space with more galaxies -- the maximum value is at \mbox{$r=0.65R_{200}^{\rm{grp}}$}, \mbox{$v=1.15v_{\rm{cir}}$}, representing the region of this phase space in which group members are most likely to be found. The red line represents the boundness criterion for galaxies \Eq{eq:bounded}; galaxies above this line are not considered group members, and so are excluded from this figure. Contours are at densities of [0.2, 0.4, 0.6, 1, 2, 4, 6] \mbox{$(R_{200}^{\rm{grp}}v_{\rm{cir}})^{-1}$}. The data in this figure, and subsequent phase space diagrams in this work, are smoothed using a 2D kernel density estimation (KDE) with an optimised bandwidth, typically $\sim0.2$ virial units.}
\label{fig:infalling_groups}
\end{figure}

\Fig{fig:infalling_groups} shows the average distribution of galaxies in phase space, for an infalling group -- a group that has just passed within $R_{200}^{\rm{clus}}$ of the cluster centre for the first time, as shown in \Fig{fig:schematic}. Similarly to our previous work \citep{haggar2021}, we produce a smoothed distribution of galaxies using a 2D kernel density estimation (KDE) with an optimised bandwidth. In the remainder of this section, we examine how this phase space changes as a group passes through a cluster.

The 1340 infalling groups that we identify represent an average of 5.2 accreted groups per cluster -- this might appear to be a small number, but it is important to note that this is not the entire accreted group population, as this only accounts for intermediate size groups (with between five and 50 members). Although they use different mass limits to this work, \citet{berrier2009} demonstrate that about half of galaxy groups contain only two or three members, and such groups are not included in our analysis. If we do include these poor groups, we find that approximately \mbox{$14\%$} of \mbox{$z=0$} cluster galaxies in our simulations were accreted as members of a group, comparable to the results from other studies presented in \Sec{sec:intro} \citep[see also][]{haggar2021}.

\subsection{Groups beyond the cluster outskirts}
\label{sec:evol_outskirts}

\begin{figure}
	\includegraphics[width=\columnwidth]{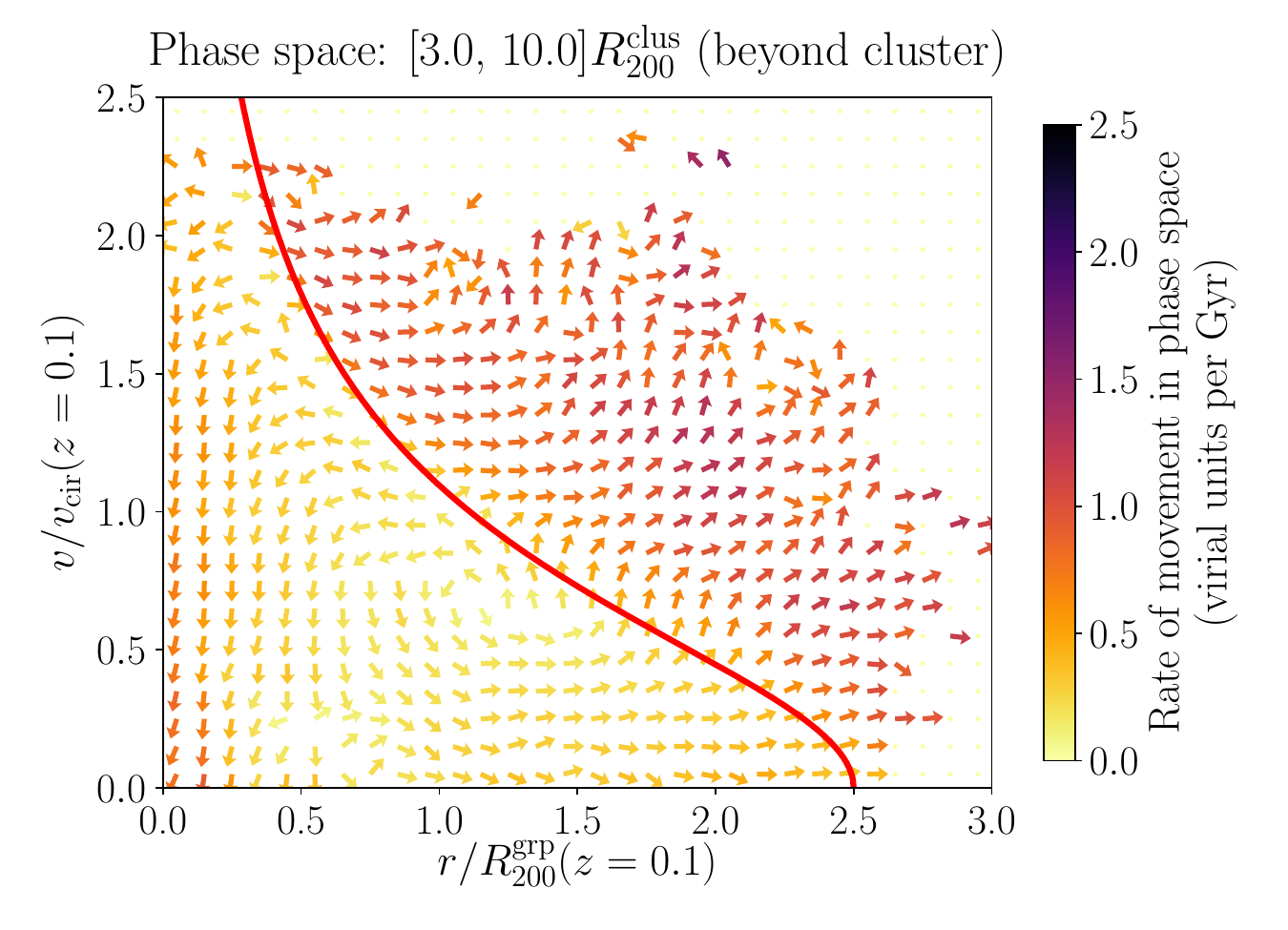}
    \caption{Motion of group members in phase space of host group, for groups located beyond the influence of a cluster. The thick red line shows boundness criterion, providing an approximate measure of galaxies that have become unbound from their group. Colours of arrows represent the rate at which galaxies are moving in this phase space, with darker arrows indicating that galaxies are moving at a greater rate in this phase space. This plot shows stacked data for 2769 groups, located between $3R_{200}^{\rm{clus}}$ and $10R_{200}^{\rm{clus}}$ from the centre of a cluster, between $z=0.1$ and $z=0$. This shows how galaxies move in the phase space of groups when the group is not affected by the external environment. All galaxies lie below the bounded line at $z=0.1$, but some move above the red line and become unbound, although many remain bound to the group. The motion of the bound galaxies follows a characteristic pattern, rather than being in random directions.}
    \label{fig:phasespace_outskirts}
\end{figure}

Before studying groups passing through clusters, we first study how this phase space changes in groups that are not under the influence of a cluster, and are located far from the cluster centre (greater than \mbox{$3R_{200}^{\rm{clus}}$} from the cluster). This can then be used as a control, showing how the distribution of galaxies changes for a group evolving secularly, as an (approximately) isolated system. \Fig{fig:phasespace_outskirts} shows the direction and rate at which galaxies in groups move around this phase space, for groups between \mbox{$3R_{200}^{\rm{clus}}$} and \mbox{$10R_{200}^{\rm{clus}}$} from the centre of a cluster, between redshifts of \mbox{$z=0.1$} and \mbox{$z=0$}. We assume that these groups are isolated, as they are sufficiently far from a cluster that they are not subject to its strongest effects. We did not study groups at greater cluster distances because the resolution of the simulations decreases outside of this distance. This figure includes only galaxies that were bound to a group (i.e. that lay below the thick red line) at $z=0.1$, which we then follow until $z=0$ (about \mbox{1.3 Gyr}).

Throughout this section, in order to study how the speeds and positions of group galaxies change, we examine the changes of these properties for bound group members, relative to \mbox{$R_{200}^{\rm{grp}}$} and \mbox{$v_{\rm{cir}}$} of their host group measured at a previous time. In \Fig{fig:phasespace_outskirts} and the following figures in \Sec{sec:evol_infall} we show how the phase space of groups changes over time. In these plots, the direction of arrows shows the average direction that galaxies in this region are moving in phase space, and darker arrows mean that the galaxies are moving across the phase space more quickly. For example, a galaxy going from \mbox{$[1.0R_{200}^{\rm{clus}}, 0.5v_{\rm{cir}}]$} to \mbox{$[2.0R_{200}^{\rm{clus}}, 1.5v_{\rm{cir}}]$} in \mbox{2 Gyr} would be represented by an arrow located at \mbox{$[1.5R_{200}^{\rm{clus}}, 1.0v_{\rm{cir}}]$}, pointing at a $45^{\circ}$ angle to the top-right, with a colour of \mbox{$\sim0.71$}. Note that the horizontal and vertical axes in \Fig{fig:phasespace_outskirts} are dimensionless, as they have been normalised to prior values of $R_{200}^{\rm{grp}}$ and $v_{\textrm{cir}}$, and so we describe the distance moved across this phase space in a given time with the term `virial units per Gyr'.

The positions and velocities of the galaxies in \Fig{fig:phasespace_outskirts} are scaled relative to \mbox{$R_{200}^{\rm{grp}}$} and \mbox{$v_{\rm{cir}}$} of each group at \mbox{$z=0.1$}. Some regions of phase space do not contain any arrows because of a lack of data, indicating that almost no galaxies were found in this region across all the groups -- for example, there are no galaxies in the top-right of \Fig{fig:phasespace_outskirts}, because they were all below the red line just a short time previously. Some galaxies are still found above the line, because they have become unbound between \mbox{$z=0.1$} and \mbox{$z=0$}. 

The phase space of these groups is not in equilibrium, and bound galaxies in the centres of these groups appear to be moving downwards on this plot (i.e. losing speed but remaining a similar distance from the group centre). This indicates that energy is being dissipated during their orbits. Dynamical friction is strongest in the group centres, and so this is likely responsible for the loss of energy during these orbits. \mbox{Fig. 2} in \citet{arthur2019} shows analogous behaviour to this for the phase space of a galaxy cluster: subhaloes move horizontally in phase space when approaching the centre of their host halo, then move sharply downwards when they are near to the halo centre, resulting in the apparent `spiral' motion of galaxies in \Fig{fig:phasespace_outskirts}. 

This trend could also potentially be explained by the destruction of some inner galaxies by mergers before they have time to leave the group centre. However, we find that this is not the case, as the majority ($82\%$) of galaxies within \mbox{$0.5R_{200}^{\rm{grp}}$} of the group centre survive to $z=0$ without merging into another halo or being heavily stripped (see \Sec{sec:fates} for further discussion of the fates of group members). Furthermore, if we remove these galaxies from our analysis, there is a negligible change in the trends in \Fig{fig:phasespace_outskirts}.

\subsection{Groups passing through clusters}
\label{sec:evol_infall}

To study groups falling into clusters, the phase space diagrams that we present are instead scaled relative to \mbox{$R_{200}^{\rm{grp}}$} and \mbox{$v_{\rm{cir}}$} of each group at $z_{\rm{infall}}$, the moment of cluster infall. We scale the positions and speeds of galaxies by these values, even in subsequent snapshots after $z_{\rm{infall}}$. This approach is not perfect, because the radius and circular velocity of a host group halo changes as the group approaches the centre of a cluster, likely due to tidal stripping. Despite this, we choose to measure these properties only at the moment of infall because, in the central regions of a large halo, the mass and radius of a subhalo are not well-defined; due to the high background density in the centre of the cluster halo, it can be challenging for a halo finder to identify the overdensity of a subhalo. Consequently, near the centre of a cluster, the mass and radius of a group ($M^{\rm{grp}}_{200}$ and $R^{\rm{grp}}_{200}$) are not reliable \citep{muldrew2011}. Scaling by the values of $R_{200}^{\rm{grp}}$ and $v_{\rm{cir}}$ at $z_{\rm{infall}}$ allows us to visualise how the absolute values of the distance and speed of galaxies relative to their groups are changing. This means that galaxies lying below the line of boundness after infall are not strictly bound to the group, but the approach still provides a good approximation.

In this section we consider groups that are entering the cluster for the first time, and so have not previously experienced a cluster potential. We also only include groups at times between their first infall, and their first apocentric passage after entering the cluster (the turnaround in their cluster orbit). It is important to note that this is not necessarily the true `first apocentre' of an orbit, as haloes are not accreted onto clusters in perfectly radial orbits. Instead, they have some tangential component to their velocity, meaning that some haloes will pass an apocentre before their entry to the cluster \citep{ghigna1998,tollet2017}. However, as the focus of this paper is on the evolution of groups after their cluster infall, we will hereafter refer to the first apocentric passage post-infall as the `first apocentre'. Finally, we do not separate groups by redshift -- for example, some of these groups have passed their first pericentre by $z=0$, but some have not and so are absent from the post-pericentre analysis.

\begin{figure*}
\includegraphics[width=\textwidth]{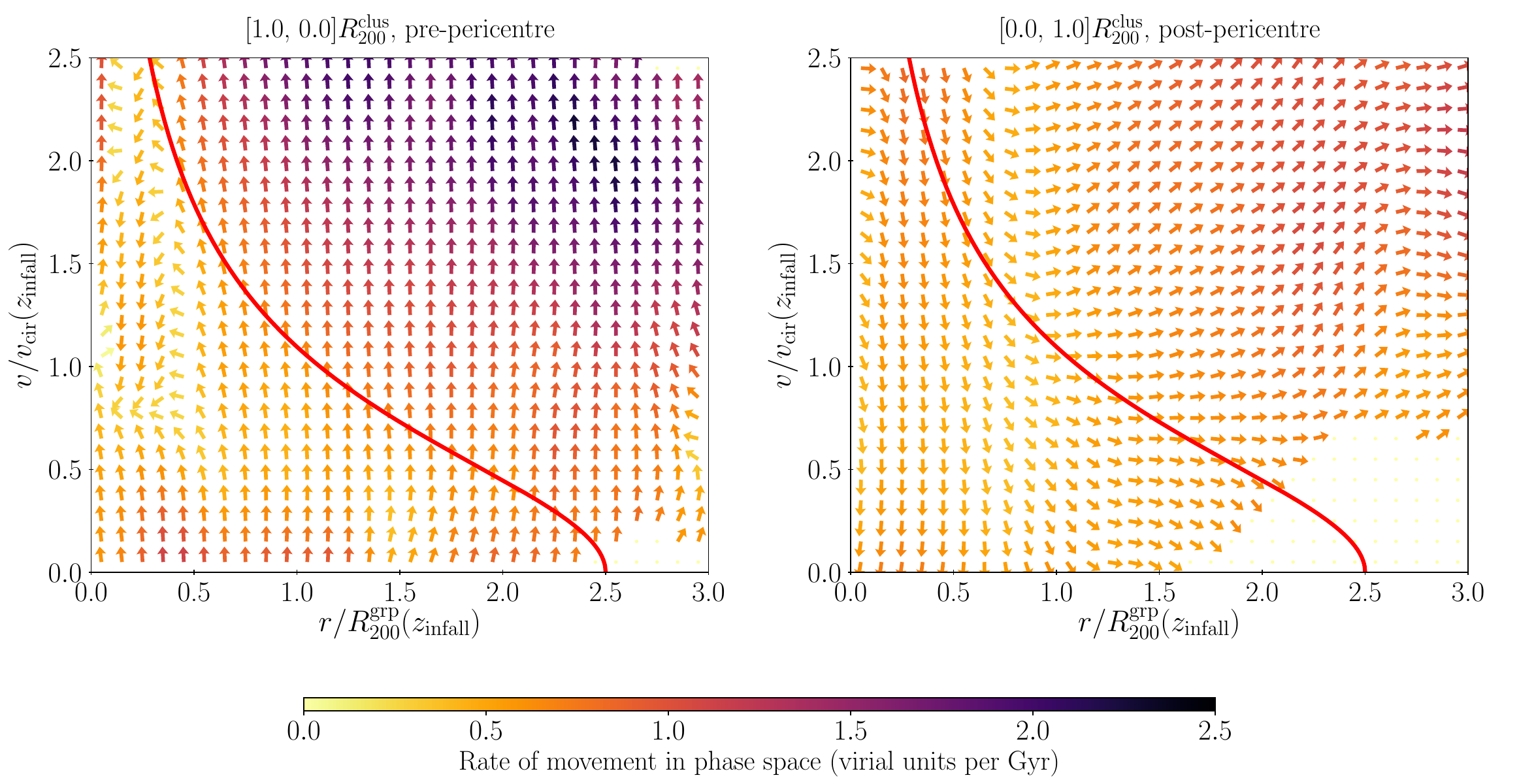}
\caption{Same as \Fig{fig:phasespace_outskirts}, but showing motion of group members in phase space of host group for two epochs, before and after pericentre. Data shown are for galaxies that are bound to groups at the moment of infall, for groups on their first passage through the cluster. The red line shows boundness criterion at the moment of infall, and so provides an approximate measure of galaxies that have become unbound from their group. Left panel shows data for groups before reaching their first pericentric passage of the cluster, moving between $R_{200}^{\rm{clus}}$ and cluster centre, and the right panel shows groups moving between the cluster centre and $R_{200}^{\rm{clus}}$, which have passed their pericentre and are now receding from the cluster, moving towards their first apocentric passage. For pre-pericentre groups, the bulk motion of the galaxies is upwards, representing an increase in their group-centric speed, but little change in the spatial separation of galaxies from their host group. In contrast, for groups that have passed pericentre and are now receding from the cluster, group members are moving approximately horizontally in phase space, increasing their distance from the group to which they were previously bound.}
\label{fig:phasespace_two-phase}
\end{figure*}

\Fig{fig:phasespace_two-phase} shows how the phase space of these groups changes as they pass through a cluster. We find that the behaviour of groups as they enter and pass through a cluster can be approximately split into two main phases, with the group dynamics changing suddenly as a group makes its closest approach to the cluster centre, as shown by the two panels in this figure. The left panel of \Fig{fig:phasespace_two-phase} shows groups on their infall, moving from the cluster outskirts towards their first pericentric passage, near to the cluster centre. Generally, the galaxies in these groups move upwards on this plot, showing an increase in their velocity relative to their host group. This data is for galaxies that are bound to groups at the moment of infall ($z_{\rm{infall}}$), but some of these move above the red line and so become unbound from their host group. Similarly to in \Fig{fig:phasespace_outskirts}, the direction of arrows shows the average direction that galaxies are moving in phase space, for galaxies in this region of phase space. It is important to note that the arrows in the left panel (`pre-pericentre') are pointing vertically upwards, with a very small horizontal component. This shows that, although these galaxies have a large change in speed, their distance to the group centre does not change very much; galaxies within $R_{200}^{\rm{grp}}$ remain within $R_{200}^{\rm{grp}}$. 

This behaviour is different for groups that have passed the pericentre of their orbit, shown in the right panel of \Fig{fig:phasespace_two-phase}. This panel shows data for groups at snapshots when they have passed pericentre, but have not yet reached their first apocentre, and so are receding from the cluster centre. Groups are also only included here at stages of their orbit when they are between the cluster centre and $R_{200}^{\rm{clus}}$, to allow us to compare the two panels in \Fig{fig:phasespace_two-phase}. Instead of increasing their velocity, most galaxies in these post-pericentre, receding groups keep a fairly constant relative velocity, and instead move horizontally on this plot, becoming spatially separated from the centre of their host group. This behaviour is stronger for galaxies that have become unbound from the group, moving above the boundness line -- these move to greater distances from the group centre, often with an accompanying slight increase in relative speed. Galaxies that are still bound to a group instead experience a drop in their relative speed, as well as an increase in separation from the group centre.

In summary, \Fig{fig:phasespace_two-phase} shows that there are two phases of evolution for a galaxy group passing through a large cluster. First, galaxies are given a kinetic energy kick, increasing their speed relative to their host group. This rapid boost in kinetic energy is manifested after the group passes pericentre, which typically occurs $\sim0.5$ Gyr after entering the cluster, by being converted into potential energy as the galaxies recede from the group centre.

\subsubsection{Tidal effects and dynamical friction}
\label{sec:tidal_friction}

In \Fig{fig:phasespace_arrows} we break down the results from \Sec{sec:evol_infall} into individual steps, separating the infalling groups into bins based on their cluster-centric distance, both before and after passing pericentre. This gives a much more detailed view of how this phase space changes over the average course of a group through a cluster. We note that each panel does not represent an identical sample of groups, as most groups will not have a snapshot in all of these radial bins, and so this data represents the evolution of all groups that are found in this radial range. If we instead select only groups that have passed through each of these bins, there is only a minimal impact on our results, but large amounts of noise are introduced due to the small number of groups. 

In each panel, the tidal radius (based on the approximations detailed in \Sec{sec:tidal_radius}) for a group in the centre of this bin is also marked, in units of the group radius at infall. The closer a group is to the cluster centre, the stronger the effect of the cluster will be, and this is demonstrated by the decrease and subsequent increase of the tidal radius as groups pass through the cluster. Interestingly, across the eight panels, the tidal radius appears to mark a transition, such that the group dynamics evolve differently within the tidal radius, compared to beyond the tidal radius. Outside the tidal radius, galaxies first experience a kinetic kick and then recede from the group centre, as detailed in the previous section. However, inside the tidal radius, galaxies generally behave in a way similar to that seen in the centres of isolated groups in \Fig{fig:phasespace_outskirts} -- they mostly move downwards on these plots, showing a decrease in speed. 

Physically, this distinction indicates how the dynamics in some regions of the group are dominated by the group itself, whilst others are dominated by the effects of the cluster. As described in \Sec{sec:evol_outskirts}, galaxies in isolated groups experience dynamical friction due to the group's halo \citep{vijayaraghavan2015}. This is particularly strong in the dense central regions of the group, where dynamical friction will cause galaxies to slow down and spiral inwards, dominating over the effect of the cluster. However, beyond the tidal radius, tidal effects from the cluster dominate this dynamical friction, meaning that the movement of galaxies in phase space is dictated by the cluster, not the group. The change in the tidal radius means that the two phases of group evolution are clearer in the outskirts of a group, as the dynamics of these regions are dominated by the cluster for much of the group's journey. Conversely, galaxies in the group centres ($r<0.5R_{200}^{\rm{grp}}$) decrease in velocity at almost all times, as they are almost always within the tidal radius. The only exception to this is in the very deepest parts of the cluster (such as in panel~(d) in \Fig{fig:phasespace_arrows}). \citet{dekel2003} showed that at the very centre of a dark matter halo, tidal forces can become fully compressive -- this could explain why all group galaxies change their orbits around their host groups, with their speeds increasing and their distances either remaining the same or decreasing.

Finally, as the change from an increase in $v$ to an increase in $r$ is dependent on the tidal radius, this switch in behaviour is not instantaneous as it might appear to be in \Fig{fig:phasespace_two-phase}. Once a group reaches a distance of approximately $0.3R_{200}^{\rm{clus}}$ beyond pericentre (panel~(f) in \Fig{fig:phasespace_arrows}), the motion of galaxies away from the group begins in the centre, and then spreads throughout the group as it once again dominates over the cluster. Eventually, for groups that are long past pericentre (panel~(h) of \Fig{fig:phasespace_arrows}), all galaxies are either decreasing in relative speed, or their speed is staying the same. All the galaxies remaining in groups at this stage are also moving away from the group centre, towards the bottom-right of the phase space, which is characteristic of galaxies approaching the apocentre of a bound orbit around a group.

To help visualise this behaviour, \App{sec:appendix_example_cluster} shows how group galaxies move around phase space for a single example group as it passes through a cluster, clearly showing the two main phases of group evolution.

\section{Groups after cluster infall}
\label{sec:before_after}

The results in \Sec{sec:evolution} show how the dynamics of galaxy groups change as they pass through a cluster. In this section, we discuss the differences in the properties of a group before and after it passes through a cluster, in order to understand how distinguishable are these two classes of groups.

\afterpage{
\begin{landscape}
\begin{figure}
\vspace{36pt}
\includegraphics[width=\columnwidth]{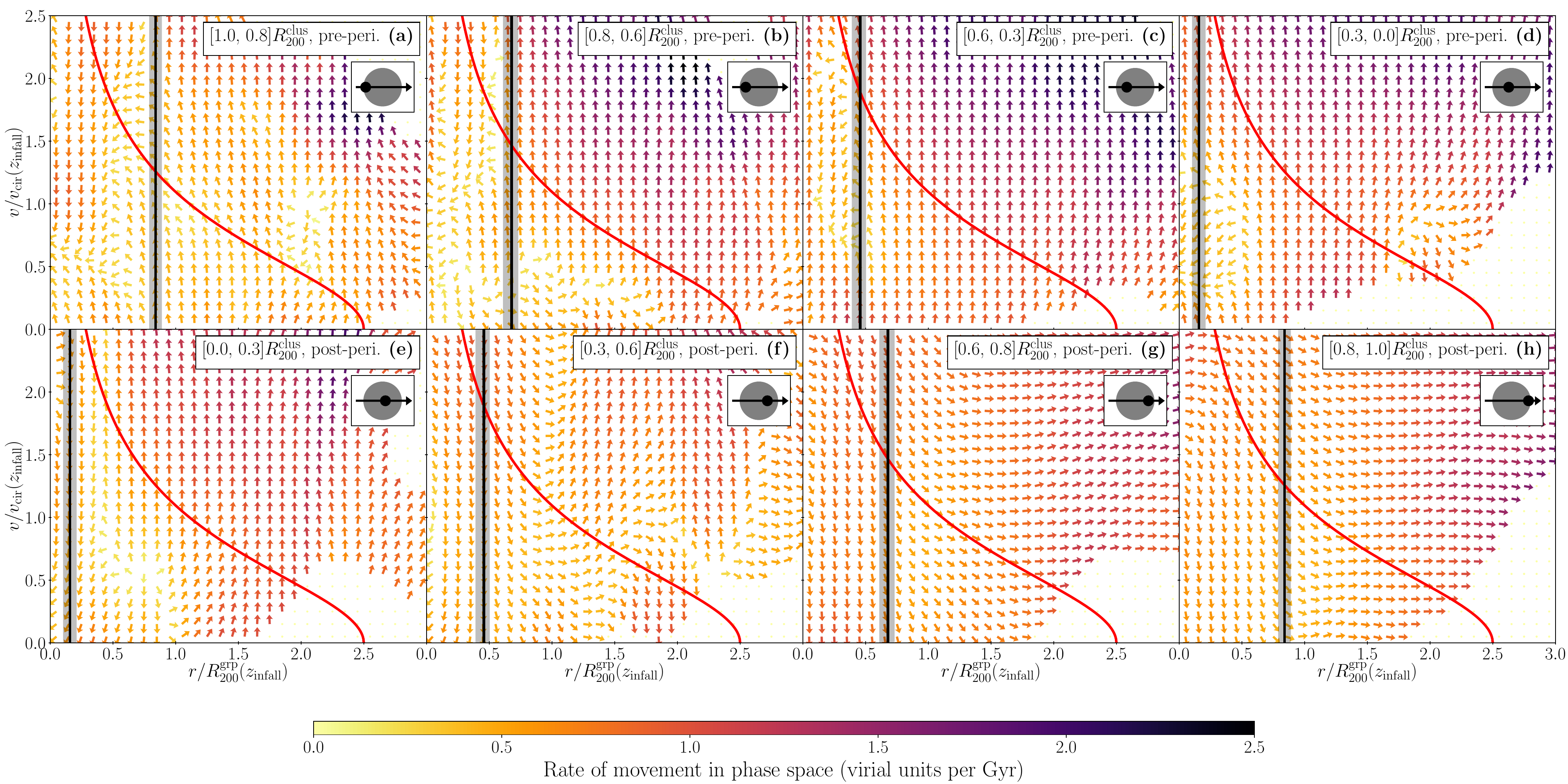}
\caption{Same as \Fig{fig:phasespace_two-phase}, but showing motion of group members in phase space of host group at multiple stages of the group's passage through a cluster. Each panel shows groups at different stages of their journey, showing the groups as they enter a cluster, approach the cluster centre, and recede from the cluster out to a distance of $R_{200}^{\rm{clus}}$. Top row (panels (a)-(d)) shows groups before reaching pericentre, binned by their distance from the cluster centre. Bottom row ((e)-(h)) shows groups after passing pericentre, before reaching their first apocentre. Data are shown for groups on their first infall only. Vertical black line on each plot represents the tidal radius for a group in this radial bin, and the shaded region represents the variation of this radius due to the $1\sigma$ spread in the concentrations of clusters and groups -- these have median halo concentrations of \mbox{$C_{\rm{c}}=3.9^{+1.5}_{-1.1}$} and \mbox{$C_{\rm{g}}=4.4^{+2.5}_{-1.8}$}, respectively. A small schematic is shown in the top-right of each panel, showing the point at which the groups (small black circle) are on their journey through the cluster halo (large grey circle) -- these schematics show a radial orbit, although most infalling group orbits also have some tangential component to their velocity. The transition between the two phases of group evolution shown in \Fig{fig:phasespace_two-phase} can be seen, as well as the corresponding change in the tidal radius discussed in \Sec{sec:tidal_friction}.}
\label{fig:phasespace_arrows}
\end{figure}
\end{landscape}}

\subsection{Orbits of galaxy groups}
\label{sec:orbits}

The data used in \Sec{sec:evolution} is for groups on their first passage through a cluster, but not all of these groups will follow the same path. Just as some galaxies that are accreted by a cluster can become `backsplash galaxies' \citep{balogh2000,gill2005,haggar2020}, some groups will pass through the cluster and exit $R_{200}^{\rm{clus}}$ again, becoming `backsplash groups' that can enter the cluster for a second time. Others will `stick' to the cluster, remaining bound and not leaving $R_{200}^{\rm{clus}}$ once they have entered. 

We find that, across the 257 clusters used in this work, most groups ($92\%$) that fall into a cluster do not leave it again. By \mbox{$z=0$}, only 42\% of the groups that enter a cluster have reached their first turnaround (apocentre) in their cluster orbit, while $58\%$ do not reach this stage. These groups do not reach apocentre for multiple reasons; either they have merged with the cluster halo before reaching apocentre rather than remain a substructure of the cluster, they have been heavily stripped by the cluster and so fall below the resolution limit before reaching apocentre, or they have simply not had time to reach apocentre by \mbox{$z=0$}. Of the groups that do reach the apocentre of their orbit, 20\% have left the cluster after entering $R_{200}^{\rm{clus}}$, while 80\% reach apocentre within $R_{200}^{\rm{clus}}$ of the cluster centre, and so remain `stuck' to the cluster potential. We hereafter refer to these as `backsplash groups' and `sticky groups', respectively. Overall, just 8\% of all infalling groups go on to leave the cluster again. Finally, 81\% of the backsplash groups in our sample later experience a second cluster infall, and 19\% are still outside of the cluster at \mbox{$z=0$}.

The paths that groups can take through a cluster can be described in terms of the distance from a group to the cluster centre at pericentre and apocentre. Interestingly, we find that the distance of a group halo from the cluster centre at pericentre is very consistent, regardless of the group's later behaviour. Groups that do not reach apocentre have a median pericentric distance of \mbox{$0.36^{+0.14}_{-0.09}R_{200}^{\rm{clus}}$} from the cluster centre, which is very similar to the pericentre of groups that do later reach apocentre: backsplash groups and sticky groups have median pericentric distances of \mbox{$0.38\pm0.13R_{200}^{\rm{clus}}$} and \mbox{$0.36^{+0.11}_{-0.08}R_{200}^{\rm{clus}}$}, respectively. This justifies our decision to normalise the figures throughout this paper by the group radius at infall, as most groups pass well within \mbox{$0.7R_{200}^{\rm{clus}}$}, where \citet{muldrew2011} showed that subhalo sizes cannot be reliably measured.

This shows that most groups take a similar trajectory into clusters, passing by the cluster centre at a similar distance. However, the subsequent orbits of these groups can vary dramatically, with groups reaching a wide range of apocentric distances, and some not being tracked to reach their apocentre at all. By definition, the post-infall apocentric cluster distances of backsplash groups and sticky groups are very different. Backsplash groups have a median apocentric distance of \mbox{$1.16^{+0.29}_{-0.12}R_{200}^{\rm{clus}}$}, and sticky groups of \mbox{$0.63^{+0.22}_{-0.16}R_{200}^{\rm{clus}}$}, which correspond to median orbital eccentricities of \mbox{$0.53\pm0.12$} and \mbox{$0.25^{+0.19}_{-0.15}$}, respectively.

\subsection{Removal of galaxies from groups}
\label{sec:removal_groups}

The sample of backsplash groups that exit a cluster and then re-enter allow us to directly compare how a single passage through a cluster permanently affects the properties of a group. Comparing the same sample of groups at the moment of first infall and second infall means that the groups are in approximately the same configuration (at a distance of $\sim{R}_{200}^{\rm{clus}}$, falling towards the cluster). 

Overall, we find that groups on a second infall contain far fewer galaxies, when compared to groups infalling for the first time. On their first infall, the median number of galaxies in these groups was $6^{+3}_{-1}$ (note that this is slightly smaller than the value of \mbox{$8^{+7}_{-3}$} quoted in \Sec{sec:evolution}, which includes groups that do not exit and re-enter the cluster). By their second infall, the median richness of these same groups is $2\pm{1}$ members. In fact, 46\% of the groups that are infalling for the second time contain only one member. Physically, these objects are not actually groups at all: a `group' with one member instead represents a single galaxy that has no other galaxies bound to it, having previously been the central galaxy in a group. This shows that, in a single passage through a cluster, almost all galaxies become unbound from groups. Often this process completely disrupts a group, resulting in no galaxies remaining bound together.

Similarly, the dark matter haloes of these groups are heavily stripped during their passage through the cluster. At first infall, the median radius, $R_{200}^{\rm{grp}}$, of a group was \mbox{$0.51^{+0.15}_{-0.10}\ h^{-1}$ Mpc}. By their second infall, these same groups had a median radius of \mbox{$0.32^{+0.10}_{-0.07}\ h^{-1}$ Mpc}. Similarly, the median mass\footnote{The infall mass of groups that later have a second infall is slightly smaller than the average mass of all groups, \mbox{$10^{13.5}\ h^{-1}M_{\odot}$} (\Sec{sec:infalling}). However, as we discuss in \Sec{sec:fates}, we still consider these to be a representative sample of all infalling groups.} of these groups, $M_{200}^{\rm{grp}}$, decreases by a factor of three in this time, from \mbox{$10^{13.2}\ h^{-1}M_{\odot}$} to \mbox{$10^{12.7}\ h^{-1}M_{\odot}$}, consistent with the decrease in the number of galaxies. This is comparable to the results from other previous studies which have found that dark matter subhaloes are heavily stripped; \citet{muldrew2011} found that a halo passing through the centre of a cluster has approximately half of its mass stripped away, and \citet{taylor2004} used semi-analytic models to show that subhaloes on orbits similar to our groups lose $>40\%$ of their mass with each pericentric passage of a cluster. Some studies find even more extreme evidence of this removal of dark matter: \citet{smith2016} used hydrodynamical simulations to show that a cluster halo can strip away $\sim80\%$ of the dark matter in galaxy-sized subhaloes.

\subsection{The fates of group galaxies}
\label{sec:fates}

We can investigate the removal of galaxies from groups further, by comparing these groups at different stages of their infall and journey through a cluster. As shown in \Sec{sec:evolution}, the speed of galaxies relative to their host group increases before they have reached pericentre of their cluster orbit, and their group-centric distance increases post-pericentre. Therefore, although groups become spatially separated after pericentre, it is not clear when the galaxies become unbound from these groups.

For backsplash groups that also have a second infall, their member galaxies are removed from their host group very quickly. Of those galaxies that are bound to a group at first cluster infall (i.e. that satisfy \Eq{eq:bounded}), $60^{+20}_{-35}\%$ are no longer bound to the group by the first pericentre, $76^{+24}_{-6}\%$ are removed by apocentre, and $89^{+11}_{-29}\%$ by the second infall into the cluster (median and $1\sigma$ spread for backsplash groups). These numbers are almost identical for backsplash groups that do not have a second infall. 

For groups that reach apocentre but do not leave the cluster (`sticky groups'), $75\pm{25}\%$ of previously bound galaxies are no longer group members at pericentre, and $73^{+27}_{-33}\%$ at apocentre. Although it appears that the number of unbound galaxies drops slightly between pericentre and apocentre, this can actually be explained by the fact that the radius (and thus mass) of a subhalo are artificially suppressed in the centre of a large halo, making more galaxies appear to be unbound. 

However, although these galaxies are no longer members of the group, this is not necessarily because they have become gravitationally unbound from their host group. In this section, we analyse the final fates of these galaxies after their group enters a cluster. To do this, we separate the galaxies' states into five categories:

\begin{itemize}
    \item Bound: galaxy is still bound to its host group, according to \Eq{eq:bounded}.
    \item Unbound: galaxy does not satisfy \Eq{eq:bounded}, and so is no longer bound to its host group.
    \item Disrupted: no descendent of a group member has been found by the halo finder, typically because its dark matter halo has been heavily stripped.
    \item Merged with group: galaxy has been absorbed by the halo of its host group, effectively merging with the Brightest Group Galaxy.
    \item Other merger: merging with another, larger object (e.g. merging with a more massive satellite galaxy). Alternatively, galaxy may be absorbed by the cluster halo, effectively merging with the Brightest Cluster Galaxy.
\end{itemize}

The `disrupted' galaxies in our sample represent a class of objects that have physical similarities. However, because of the nature of the simulations and tree-builder that we use in this work, the branches of their merger trees are cut off prematurely, meaning that they appear to have no descendent halo in the simulations and so their final fate cannot be determined. Before their branches end, the dark matter masses of these galaxies are changing rapidly -- in their final ten snapshots before they are removed from the merger tree, $76\%$ of these galaxies experience at least one drop of $>30\%$ in their halo mass between two snapshots \mbox{($\sim0.3$ Gyr)}, and $37\%$ have a measured drop of $>40\%$. However, {\sc{mergertree}} does not allow for the dark matter mass of an object to change by more than a factor of two between snapshots (see \Sec{sec:tree} for an explanation of this). Consequently, if a galaxy's dark matter halo mass drops by $>50\%$ between snapshots, this change will not be recorded, no descendent for the halo will be added to the catalogue, and this branch in the merger tree will end. Despite this heavy stripping of dark matter, very few of the disrupted galaxies violate the mass limits that are imposed in \Sec{sec:tree}; if we remove these mass limits, the median final mass of these galaxies before their merger tree ends is \mbox{${\rm{log}}_{10}(M_{200}/h^{-1}M_{\odot})=11.3^{+0.7}_{-0.6}$}, with a median stellar mass of \mbox{${\rm{log}}_{10}(M_{\rm{star}}/h^{-1}M_{\odot})=10.4^{+0.5}_{-0.4}$}, and a ratio between these of \mbox{$0.14^{+0.12}_{-0.08}$}. Consequently, few of these galaxies are removed from the merger trees due to violating these imposed mass limits.

\begin{figure}
\includegraphics[width=\columnwidth]{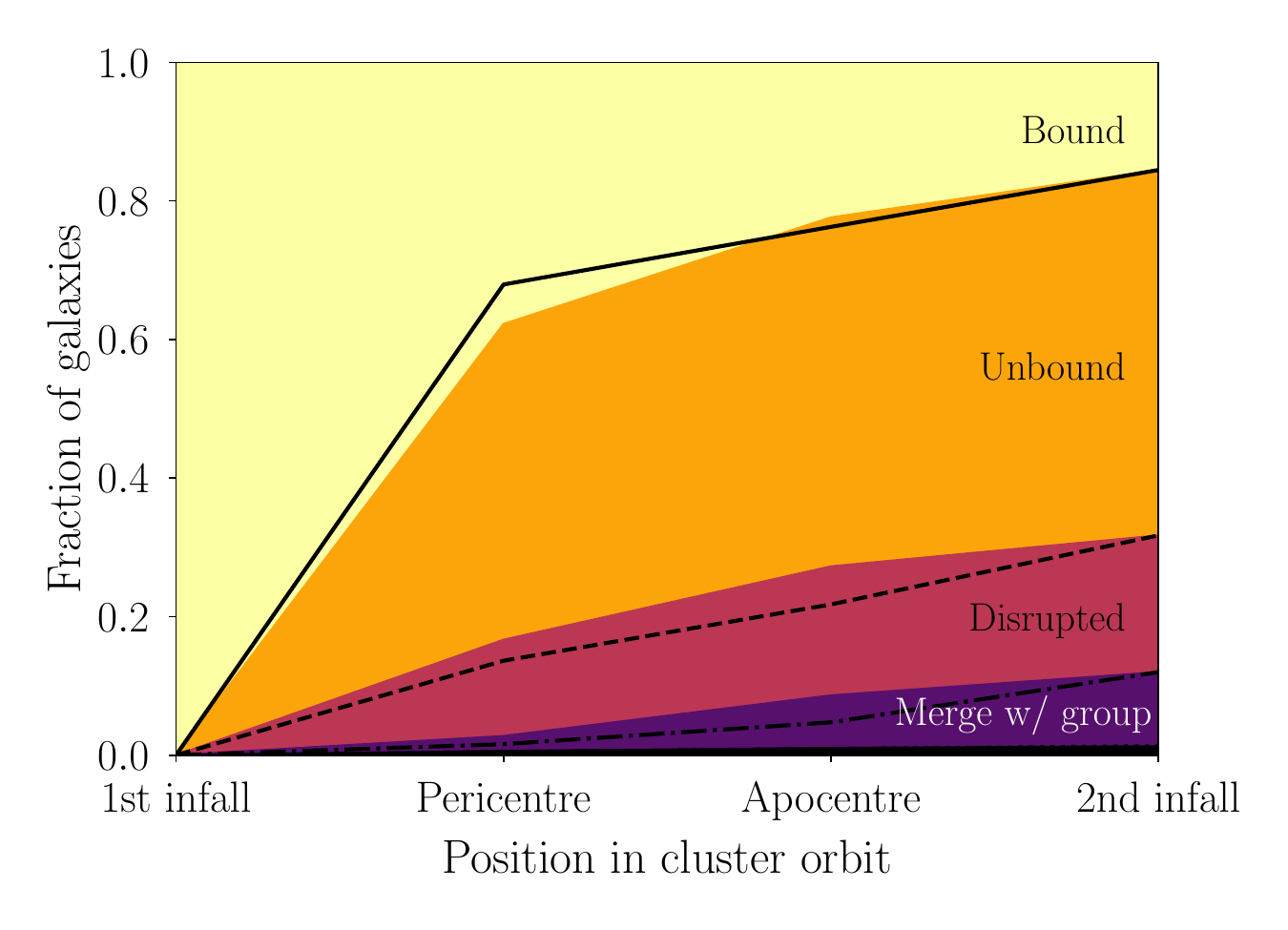}
\caption{Status of galaxies that were bound to groups at cluster infall, as their host group passes through a cluster and begins its second infall. These data are averaged across all groups that become backsplash groups, and experience a second infall. All galaxies are bound at first infall, by definition. Areas representing galaxies that have become bound, unbound, disrupted, or merged with the group halo are labelled. The small, black region represents other mergers, which is unlabelled for clarity. Solid/dashed/dot-dashed/dotted lines show the boundaries between these regions for all groups that reach this stage of their orbit, regardless of whether they go on to reach apocentre or have a second infall.}
\label{fig:galaxy_fates}
\end{figure}

\Fig{fig:galaxy_fates} shows the status of group member galaxies as their host group passes through a cluster. This data is averaged across all groups that become backsplash groups and then have a second cluster infall, meaning that we have data for their entire passage through a cluster. Overlaid as solid, dashed, dot-dashed and dotted lines are the boundaries between the coloured regions when all groups are included. For example, this indicates the states of galaxies at pericentre for all groups that reach their first pericentre, regardless of what subsequently happens to the group. Similarly, the apocentre data shows the fates of all galaxies in groups at apocentre, whether or not this apocentre is outside of the cluster. This data closely follows the data for groups that have a second infall, showing that these second infallers are representative of the entire group sample. We therefore only discuss these groups that later have a second infall, allowing us to make comparisons of the same sample of groups at different stages of their orbit.

As stated above, only approximately $40\%$ of group members are still members of the group at the pericentric passage of the cluster centre (note that here we use the mean behaviour of each group, as opposed to the median used earlier on in this section, and so the quantities differ slightly). However, of the $61\%$ of galaxies that are no longer group members at pericentre, only $45\%$ have become unbound from their host group, while $16\%$ have experienced one of the other fates described above. As these groups exit the cluster and re-enter, the number of galaxies becoming unbound increases slightly (to $53\%$), but the number of galaxies leaving the group for another reason doubles, to $32\%$, showing that these other processes are more important after a group's initial infall. 

It is also important to note that these four stages in the group orbit -- infall, pericentre, apocentre and second infall -- are not equally spaced in time. For the groups shown in \Fig{fig:galaxy_fates} (backsplash groups with a second infall), pericentre, apocentre and the second infall occur an average of \mbox{$0.5\pm0.2$ Gyr}, \mbox{$2.6^{+0.4}_{-0.7}$ Gyr}, and \mbox{$3.5^{+1.2}_{-1.0}$ Gyr} after the first infall, respectively. Consequently, not only do most of the unbound galaxies leave the group between infall and pericentre, but this process takes place in just \mbox{$\sim0.5$ Gyr}, compared to the \mbox{$\sim2$ Gyr} between pericentre and apocentre. We note that these timescales are redshift dependent: the time for a galaxy entering a cluster to reach pericentre at $z=0$ can typically range from $1-2$ Gyr \citep[see Fig. B1 in][for further details]{tollet2017}, but the time taken decreases at higher redshifts. Our method consequently returns an average infall-to-pericentre time of less than $1$ Gyr, because we stack data from groups at numerous different redshifts \citep[for some additional discussion of cluster crossing times, see][]{contrerassantos2022}.

\begin{figure}
\includegraphics[width=\columnwidth]{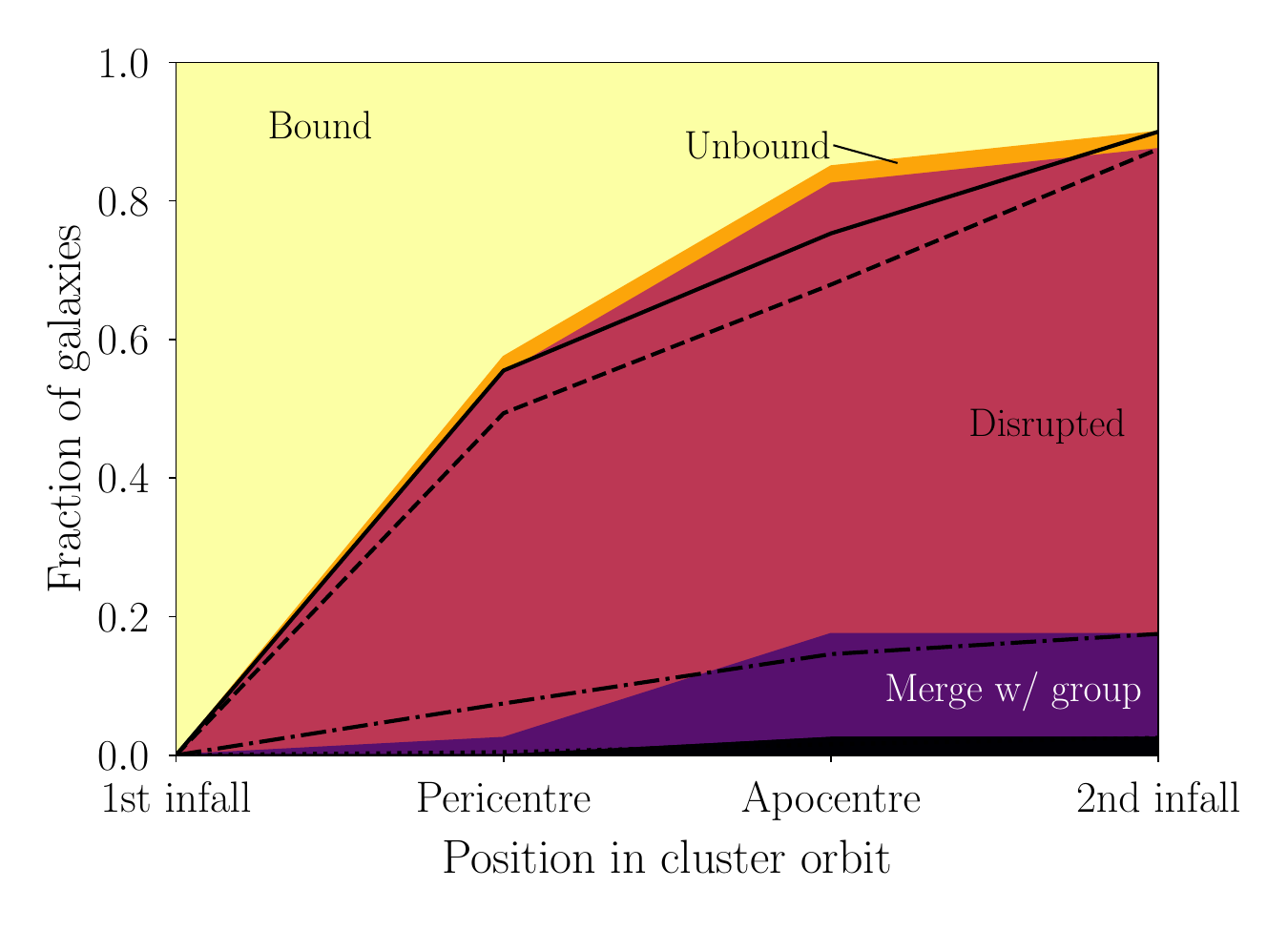}
\caption{Same as \Fig{fig:galaxy_fates}, but for slow-moving galaxies in the centre of groups ($r<0.5R_{200}^{\rm{grp}}$, $v<0.5v_{\rm{cir}}$). Again, the small `other mergers' region is unlabelled for clarity. Galaxies in this region are much more likely to become heavily disrupted and have an incomplete merger tree, although a substantial fraction merge with the group halo, mostly between pericentre and apocentre.}
\label{fig:galaxy_fates_inner}
\end{figure}

\begin{figure}
\includegraphics[width=\columnwidth]{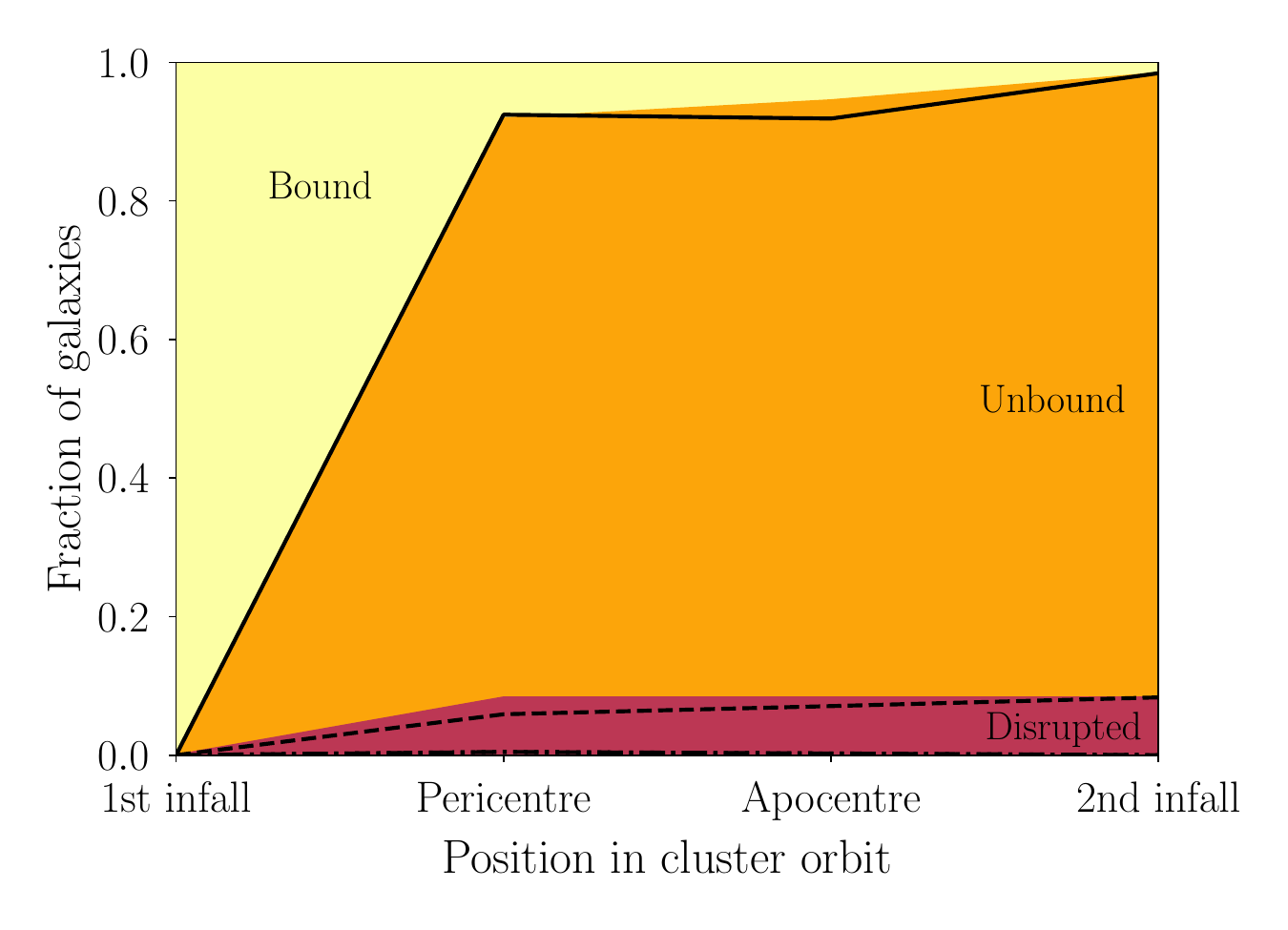}
\caption{Same as \Fig{fig:galaxy_fates}, but for galaxies in the outskirts of groups \mbox{($r>0.8R_{200}^{\rm{grp}}$)}. The `group mergers' and `other mergers' regions are unlabelled for clarity. Galaxies in the outskirts of the groups are highly likely to become unbound from their host group, which usually happens between infall and pericentric passage.}
\label{fig:galaxy_fates_outer}
\end{figure}

\Fig{fig:galaxy_fates} represents all group members at infall, however \Fig{fig:phasespace_arrows} shows that galaxies in different regions of the group phase space will experience different processes, and so the likelihood of each outcome is not the same for all galaxies in a group. Accordingly, we also find that the evolution and fates of group galaxies is strongly dependent on their position within the phase space of their host group. \Fig{fig:galaxy_fates_inner} and \Fig{fig:galaxy_fates_outer} show the evolution of members of groups that pass through and re-enter a cluster, in the bottom-left\footnote{This selection specifically examines slow-moving galaxies near the group centre, as fast-moving galaxies near the group centre exhibit different behaviour. We elaborate on this in \Sec{sec:fates_phasespace}.} \mbox{($r<0.5R_{200}^{\rm{grp}}$}, \mbox{$v<0.5v_{\rm{cir}}$)} and bottom-right \mbox{($r>0.8R_{200}^{\rm{grp}}$)} of the phase space shown in \Fig{fig:infalling_groups}. These represent the slow-moving galaxies deep within the group's potential well, and loosely-bound galaxies in the group outskirts, respectively.

Clearly, galaxies in the central (\Fig{fig:galaxy_fates_inner}) and outer (\Fig{fig:galaxy_fates_outer}) regions of a group have vastly different evolutionary histories. Slow-moving galaxies in the centres of groups almost never become unbound from the group -- instead, the majority of them are disrupted by the time the group re-enters the cluster, although a sizeable fraction ($17\%$) of them merge with the group halo. Dynamical friction likely plays a role in this, by causing these galaxies to spiral in towards the group centre, making them likely to merge with their host group's halo. This is in contrast to the outskirts of the groups, where the vast majority of group members become unbound from the group almost immediately after the group enters the cluster, and only a small fraction are heavily disrupted. In both cases, the black lines on the figures show that galaxies in other infalling groups experience similar evolution, although slightly more galaxies are disrupted in groups that become backsplash groups.

\subsubsection{Galaxy fates across group phase space}
\label{sec:fates_phasespace}

\begin{figure*}
\includegraphics[width=\textwidth]{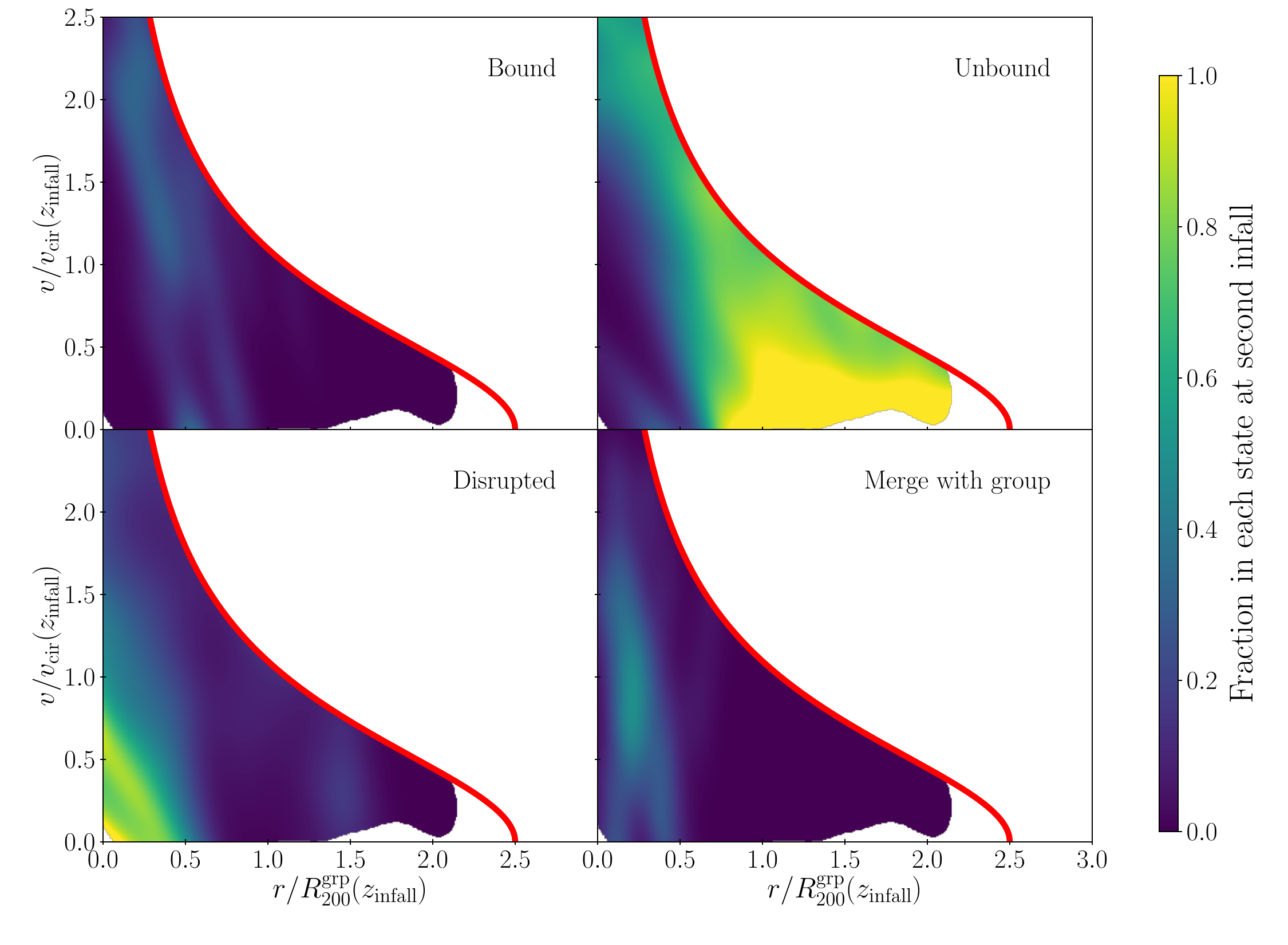}
\caption{Of the galaxies that are bound to a group at its first infall, each panel shows the fraction of these in each state at the moment of second infall. Phase space is defined by the position/speeds of the galaxies at the first infall. Top-left panel shows the fraction of galaxies that remain bound to the group. Top-right panel shows the fraction that become unbound from the group. Bottom-left shows the fraction that are `disrupted'. Bottom-right shows the fraction that merge with the group halo. Lighter colours represent regions of the phase space with a greater number of galaxies. White regions either represent the `unbound' region, or regions where the number of galaxies is very low.}
\label{fig:phasespace_difference}
\end{figure*}

Finally, we can take a more general approach to \Sec{sec:fates} by looking at the phase space of the infalling groups, to determine the typical fates of galaxies at the second cluster infall, as a function of their initial position in this phase space. \Fig{fig:phasespace_difference} shows how common different outcomes are for group members, as a function of their relative position and speed at cluster infall; this is in effect a generalisation of \Fig{fig:galaxy_fates_inner} and \Fig{fig:galaxy_fates_outer}. For example, in the bottom-left region of the phase space, there is a high density of `disrupted' galaxies, showing that galaxies here during infall later became disrupted, in agreement with \Fig{fig:galaxy_fates_inner}.

The top-left and top-right panels of \Fig{fig:phasespace_difference} show a substantial decrease in the number of galaxies that remain bound to a group outside of \mbox{$r\sim0.7R_{200}^{\rm{grp}}$} from the group centre. This indicates that, for almost all groups, virtually all galaxies outside of this radius are removed. Similarly to in \Sec{sec:evol_infall}, the tidal stripping of groups can explain this sharp cut. According to \Eq{eq:tidalrad_final}, a tidal radius of \mbox{$0.7R_{200}^{\rm{grp}}$} corresponds to a group that is approximately \mbox{$0.7R_{200}^{\rm{clus}}$} from the cluster centre. This distance is the maximum typical pericentric distance that we find for groups in our sample -- almost all groups ($95\%$) have a pericentric passage of \mbox{$r\leq0.7R_{200}^{\rm{clus}}$}. Consequently, almost all groups will have had a tidal radius of \mbox{$R_{\textrm{t}}=0.7R_{200}^{\rm{grp}}$} at some point in their orbit, but not all groups will have experienced a tidal radius less than this. This explains why some galaxies remain in the groups within $0.7R_{200}^{\rm{grp}}$, but none remain beyond this distance. 

Generally, only galaxies near to the group centre with high velocities remain as bound group members. These are on longer, eccentric orbits -- galaxies with lower velocities spend more time nearer the group centre, and so are more likely to merge with the group, or to be disrupted. Furthermore, the bottom-left and bottom-right panels show that the disrupted galaxies inhabit different parts of phase space, compared to those that later merge with the group halo. Disrupted galaxies have large amounts of their dark matter stripped in a short period of time: for two-thirds of these galaxies, in the snapshot immediately after they are `disrupted', more than $50\%$ of their dark matter particles appear either in the halo of their host group or (less often) their host cluster. This disruption by a larger halo is similar to how galaxy harassment can occur in clusters \citep{moore1996_harassment}. However, the galaxies in the centre of these disrupted haloes do not immediately become associated with the group halo -- if this were the case, these objects would be tagged as merging with the group halo, which they are not. This implies that a tidal disruption is occurring, in which large amounts of material are removed from the galaxy, forming a substructure in the group such as a tidal stream. This substructure will most likely merge with the group halo at some later time \citep{moore1998}, effectively making this process a merger with the group halo, but over a longer time period.

Disruption is more likely for galaxies in the centres of groups that are slow-moving at the moment of infall, while galaxies with greater speeds are somewhat more likely to merge with the group halo. One explanation for this lies in the left panel of \Fig{fig:phasespace_two-phase}, showing pre-pericentre groups. Before a group reaches pericentre, galaxies in the group centre with high speeds move downwards in phase space, indicating that their speed is decreasing due to dynamical friction, and they are slowly spiralling into the group centre where they merge. Slow-moving galaxies are instead moving upwards on this plot, indicating that they are experiencing strong, accelerating forces that can disrupt their structure. Additionally, high-speed galaxies are on radial, eccentric orbits, meaning that they pass the group centre infrequently. In contrast, a low speed and low group-centric distance indicates that a galaxy is on a small, circular orbit, and so will be able to make multiple orbits of the group in a short period of time, providing more opportunities for heavy stripping by the central group galaxy.

\begin{figure}
\includegraphics[width=\columnwidth]{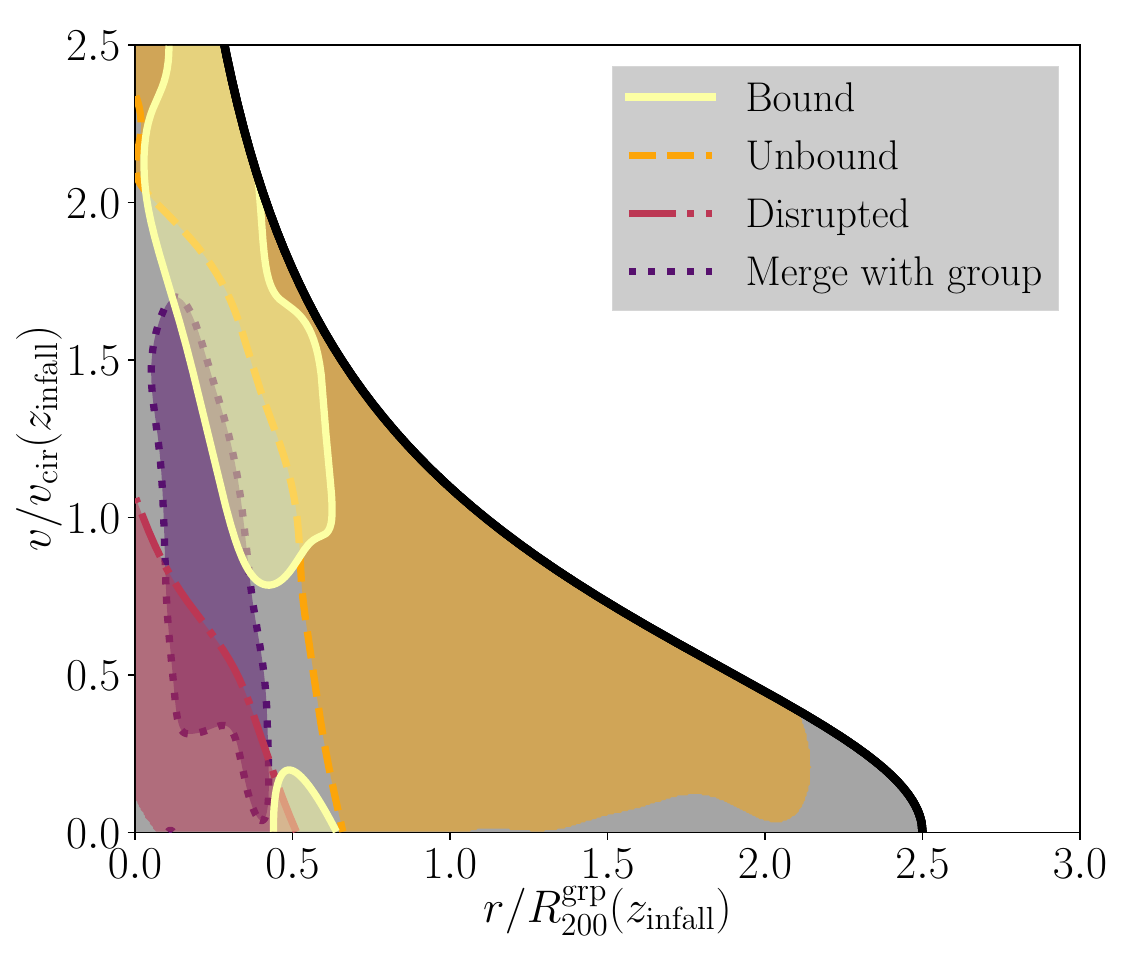}
\caption{Overlay of four panels from \Fig{fig:phasespace_difference}, to aid with comparison of different phase space distributions, for galaxies bound to groups at cluster infall. One highlighted region is shown for each of the four galaxy fates at the time of second infall. These show the region from which each class of galaxy most commonly originated -- that is, where were they previously found at the moment of first infall. Contour surrounding each region is placed at a value equal to half of the maximum, from each panel in \Fig{fig:phasespace_difference}. Grey regions either represent areas of phase space that contain few galaxies, or are not an important producer of any of these four classes of galaxies. From this, it is clear that galaxies in different regions of phase space later experience different environments, and different evolutionary processes.}
\label{fig:phasespace_contour}
\end{figure}

To more clearly show the differences between these classes of galaxies (those that are bound to the group, unbound, disrupted, or have merged with the group at second infall), we combine the four panels from \Fig{fig:phasespace_difference} into a single figure, \Fig{fig:phasespace_contour}. The contours in this show, for galaxies in each class, where in phase space they were located at the moment of infall. It is important to note that each contour is located at half of the maximum value for that class, and so they are not scaled in the same way as some galaxy fates are more common than others. Consequently, this plot does not show what outcome is most likely for galaxies in each part of phase space. For example, many more galaxies become unbound than remain bound to a group, so those in the top-left of this diagram are far more likely to become unbound than remain bound -- this is more apparent when we compare the top two panels of \Fig{fig:phasespace_difference}.

Instead, \Fig{fig:phasespace_contour} allows us to take a single class of galaxies (say, those that are disrupted by second infall), and see from where they originated (in this case, the low-velocity, inner regions of the group). Some regions are the source of multiple classes of galaxies, while some are the source of only one. For instance, group members that are later either bound or unbound can originate from the the low-$r$, high-$v$ region of phase space, but only unbound galaxies originate from the high-$r$, low-$v$ region. These results from \Fig{fig:phasespace_difference} and \Fig{fig:phasespace_contour} broadly agree with the findings of \citet{choquechallapa2019}, who use dark matter-only simulations to study the fates of galaxies in infalling groups as a function of their position in group phase space. Among other results, they show that outside of \mbox{$r\sim0.8R_{200}^{\rm{grp}}$} there is a sharp increase in the fraction of members becoming unbound from their host group, and that galaxies lying near to the boundness line are more likely to be stripped from their groups.

\subsubsection{Observational analogues}
\label{sec:obs}

This work focuses on simulations of groups and clusters, but these simulations can be used to inform and interpret future observational studies. 

It might appear that this preferential removal of outer group members could lead to the formation of very dense, centrally-concentrated groups such as Hickson Compact Groups \citep{hickson1982} in and around clusters. As \Fig{fig:phasespace_arrows} shows though, the galaxies that remain bound to a group do not remain in the same region of phase space. Instead, the remaining galaxies are redistributed throughout the group until they follow a similar distribution to that shown in \Fig{fig:infalling_groups}. These group remnants are no more centrally concentrated than the infalling groups.

Most importantly, \Fig{fig:galaxy_fates} and the top-left panel of \Fig{fig:phasespace_difference} show that, of the galaxies that are bound to a group when it approaches a cluster, almost none are still bound to a group after just a single crossing of the cluster ($\sim2$ Gyr later). Typically, only a very small number of galaxies remain in a group, and so the remnant `groups' are usually either single galaxies, or galaxy binaries. Groups with five or more members are extremely unlikely to have previously experienced a cluster environment; across all of our simulations, less than $1\%$ of such groups entering a cluster after $z=0.1$ have previously passed through a cluster. Because of this, galaxy groups nearby to a cluster (i.e. in the cluster outskirts, just outside of $R_{200}^{\rm{clus}}$) are very unlikely to contain backsplash galaxies, which typically make up about $50\%$ of the galaxies surrounding a cluster \citep{gill2005,haggar2020}. Instead, these groups represent a population of galaxies that are unprocessed by their host clusters, but have experienced group effects in their past. This may also partly explain why unrelaxed clusters, which contain more substructure and hence more galaxy groups, are typically surrounded by fewer backsplash galaxies than relaxed clusters \citep{haggar2020}.

The fact that galaxy groups observed nearby to clusters are very likely to be on their first approach to the cluster has important implications for observational studies of galaxy evolution and environmental pre-processing. Additionally, we can infer greater detail about the histories of the galaxies in these groups. For example, cluster galaxies that are currently not in groups are unlikely to have previously experienced the dense, central regions of a group, as galaxies in group centres are much more likely to remain in their groups. Similarly, galaxies associated with groups inside clusters have almost certainly previously passed through the group centre, even if they now reside in the group outskirts. This means that they will have experienced the most extreme environmental impacts of the group. \citet{hester2006} showed that, in groups of a similar size to those used in this work \mbox{($M_{200}^{\rm{grp}}=10^{13}M_{\odot}$)}, a disk galaxy with a dark matter mass of \mbox{$10^{11}M_{\odot}$} at \mbox{$r=0.75R_{200}^{\rm{grp}}$} will have $\sim20\%$ of its disk gas removed, but if this galaxy passes within \mbox{$r=0.25R_{200}^{\rm{grp}}$} of the group centre, it can lose approximately $90\%$ of its gas. They attribute this to the stronger ram pressure stripping that takes place in group centres.

\section{Conclusions}
\label{sec:conclusions}

In this work we use hydrodynamical simulations to study the evolution of intermediate sized galaxy groups (five to 50 members with stellar masses $M_{\rm{star}}\geq10^{9.5}M_{\odot}$) in the vicinity of large galaxy clusters, and specifically from the time after the groups pass within $R_{200}$ of the cluster. We begin by studying the positions and speeds of galaxies relative to their host group in order to characterise how this `phase space' of the group changes over time, before studying the fates of group members after the passage of their group through a cluster. Our findings are summarised below.

\begin{itemize}
\item On entering a cluster, galaxy groups typically pass within $0.6R_{200}^{\rm{clus}}$ of the cluster centre. Most of these groups remain permanently bound to the cluster, although a small fraction ($\sim10\%$) reach an apocentric distance outside of the cluster's radius, $R_{200}^{\rm{clus}}$.
\item The dynamics of these groups change in two phases. First, the member galaxies increase their speeds relative to the group centre, often becoming gravitationally unbound. Then, after the group passes the pericentre of its cluster orbit (which typically occurs after $\sim0.5$ Gyr in the cluster), the distances of galaxies from the group centre increases.
\item The majority of galaxies bound to a group at its first cluster infall are no longer in the group after a full passage through the cluster. Many of these galaxies become either unbound from the group, heavily stripped, or merge with the Brightest Group Galaxy, and the fate of a galaxy depends strongly on its location within the group at the infall time. 
\item Consequently, the overwhelming majority ($>99\%$) of groups that enter a cluster are doing so for the first time in their histories. In observations, groups that are seen just outside of a cluster are very unlikely to have previously experienced a cluster environment.
\end{itemize}

Although the composition and structure of simulated galaxy groups is dependent on the physical models that are used, the results from this work still allow us to make conclusions about groups that can be applied to observational work. Groups that are observed nearby to clusters are almost certainly recent infallers, particularly groups with low velocity dispersions, as galaxies in groups become gravitationally unbound almost immediately after entering a cluster. Furthermore, any galaxies that are observed in a group that is inside a cluster will have previously passed through the group centre, and so will be severely stripped by the tidal forces and ram pressure of their host group.

In addition to the approach taken in this paper, which draws conclusions on galaxy groups that can be applied observationally, work remains to be done on the dynamics of these groups. In future work we plan to study the dynamics of these groups in greater detail. For example, the binding energy-angular momentum phase space, and the orbital parameters of galaxies, can tell us about the anisotropy of group members' orbits \citep[for example]{wojtak2008,lotz2019}, which can in turn be used to describe how virialised is a group. In our future work we will study the time evolution of these dynamical parameters. 

The work in this paper further strengthens the case that galaxy groups provide a unique way to study galaxy evolution, and particularly pre-processing. As they are almost all first-infallers, groups in the outskirts of clusters will have experienced no cluster processing, and will have a very low contamination by backsplash galaxies. Consequently, pre-processing effects will dominate over any effects from the cluster in these structures, and so studying these objects in more detail will allow us to further disentangle the environmental effects of clusters, and the effects of other cosmic environments. Processes such as gas removal and morphological changes in these galaxies will exclusively have occurred pre-infall in groups or cosmic filaments, and so ultimately the properties of galaxies in groups will help inform us of the role that environment plays in driving galaxy evolution.

\section*{Acknowledgements}

We thank the referee, Gary Mamon, for his helpful and thorough comments, which have helped to improve the quality of this paper.

This work has been made possible by \threehun\ collaboration\footnote{\url{https://www.the300-project.org}}. This work has received financial support from the European Union's Horizon 2020 Research and Innovation programme under the Marie Sk\l{}odowskaw-Curie grant agreement number 734374, i.e. the LACEGAL project\footnote{\url{https://cordis.europa.eu/project/rcn/207630\_en.html}}. \threehun\ simulations used in this paper have been performed in the MareNostrum Supercomputer at the Barcelona Supercomputing Center, thanks to CPU time granted by the Red Espa\~nola de Supercomputaci\'on. For the purpose of open access, the author has applied a creative commons attribution (CC BY) to any author accepted manuscript version arising.

RH acknowledges support from STFC through a studentship. He also thanks Andrew Benson for useful suggestions relating to the tidal radius of groups, and Liza Sazonova for extremely helpful discussions relating to merger trees. AK is supported by the Ministerio de Ciencia, Innovaci\'{o}n y Universidades (MICIU/FEDER) under research grant PGC2018-094975-C21.

This work makes use of the {\sc{SciPy}} \citep{virtanen2020}, {\sc{NumPy}} \citep{vanderwalt2011}, {\sc{Matplotlib}} \citep{hunter2007}, {\sc{SymPy}} \citep{meurer2017} and {\sc{pandas}} \citep{mckinney2010} packages for {\sc{Python}}\footnote{\url{https://www.python.org}}.

The authors contributed to this paper in the following ways: RH, UK, MEG and FRP formed the core team. RH analysed the data, produced the plots and wrote the paper along with UK, MEG and FRP. AK produced the halo catalogues and merger trees. GY supplied \threehun\ simulation data. WC assisted with interpreting the time evolution of group members and group properties, particularly in \Sec{sec:fates}. All authors had the opportunity to comment on the paper.

\section*{Data availability}

The data underlying this work has been provided by \threehun\ collaboration. The data may be shared on reasonable request to the corresponding author, with the permission of the collaboration.



\bibliographystyle{mnras}
\bibliography{main} 



\appendix

\section{Evolution of a single galaxy group in phase space}
\label{sec:appendix_example_cluster}

\Fig{fig:example_cluster_0} shows an example of one galaxy group entering, passing through, and then re-entering a cluster, as a demonstration of the process discussed throughout this work. The right column shows the dark matter halo of the cluster, represented by the grey circle, and the paths of the galaxies in a group as the group passes through the cluster. The system is rotated so that the path of the group is in the plane of the page. 

In the left column, the changing position of each group member in phase space is shown (i.e. its changing position and speed relative to the host group). Each line in phase space represents the path taken by one galaxy through phase space, as the galaxy has followed the path through the cluster shown in the right column. Six timesteps are shown altogether, from top to bottom. The two phases of dynamical evolution shown in \Fig{fig:phasespace_two-phase} and \Fig{fig:phasespace_arrows} can be seen in the evolving phase space of this group, as the galaxies move upwards in phase space as the group approaches its pericentric passage, and then from left to right after this. 

For clarity, schematics are shown in the bottom-right of each panel, similarly to those used in \Fig{fig:phasespace_arrows}, to show where the group is along its orbit through the cluster. From top to bottom, the six timesteps show the group immediately after its first infall, one snapshot after pericentric passage, shortly before exiting the cluster, shortly after exiting the cluster, at apocentre, and immediately after its second cluster infall.

The right panels show the overall behaviour of groups that we find throughout this work -- a relatively compact group of galaxies remains coherent for a short period after entering a cluster, but then becomes heavily disrupted, and the galaxies are separated from each other to large distances (shown in the final panel).

\begin{figure*}
\includegraphics[width=0.75\textwidth]{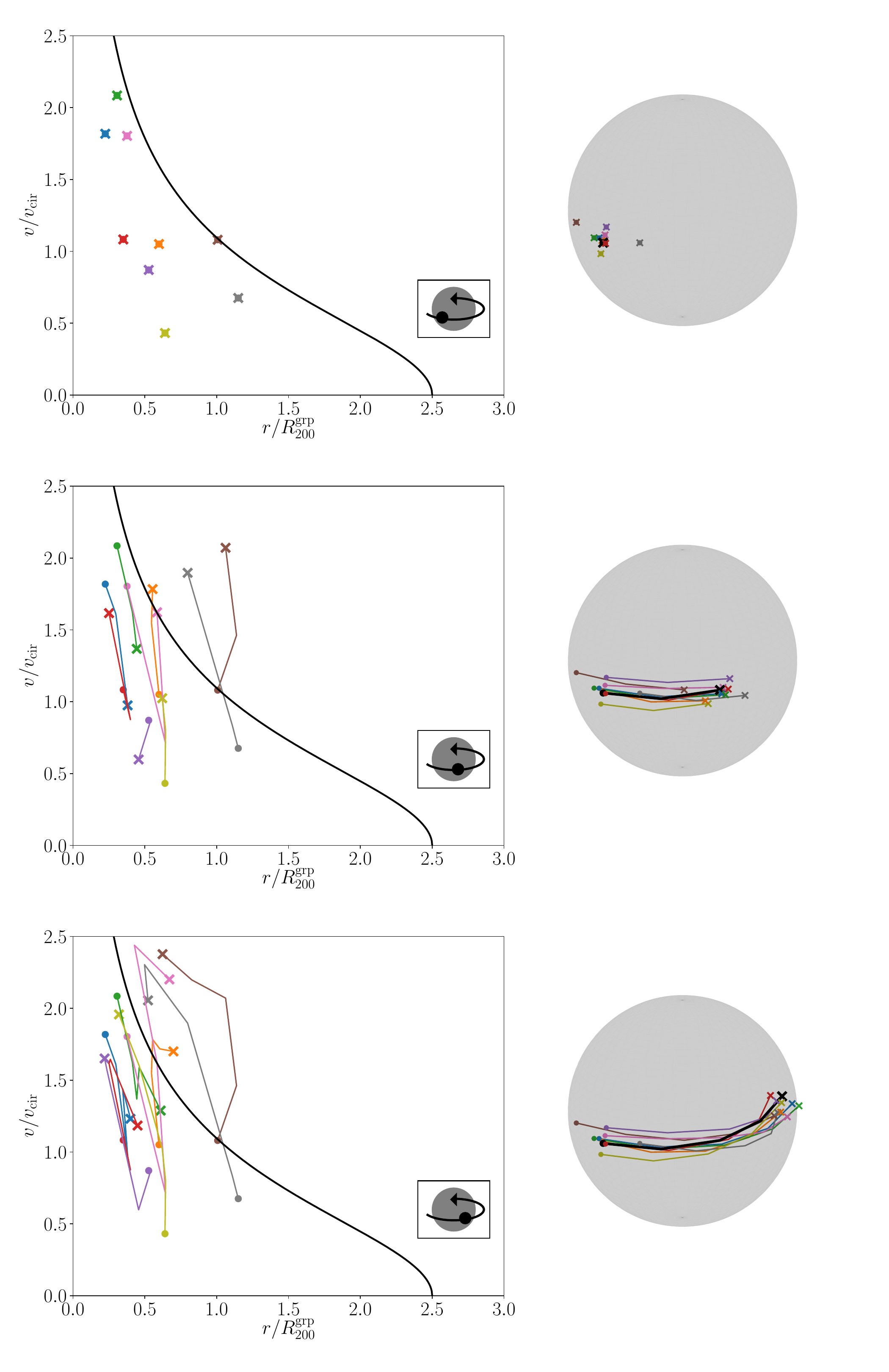}
\caption{Phase space evolution of one group as it enters a cluster, passes through, leaves the cluster and then re-enters, showing the initial increase in $v$, and subsequent increase in $r$, of the member galaxies. Right panels show the cluster halo (large grey circle), and a 2D projection of the paths taken through it by the group halo (thick black line) and the galaxies bound to this group at infall (thin coloured lines). Left panels show the corresponding paths taken by these galaxies in the phase space diagram used in \Sec{sec:evolution}, from infall to the current snapshot. For clarity, the positions of galaxies at infall and at the present time are represented by dots and crosses respectively. Six snapshots are shown altogether, with time increasing from top to bottom. Schematics in the bottom-right of the left panels show where the group is on its cluster orbit -- for instance, the fifth timestep shows the group at apocentre. The first and last panels are separated by 10 snapshots, covering \mbox{$\sim3.2$ Gyr} between \mbox{$z=0.25$} and \mbox{$z=0$} {\textit{(continued on next page)}}.}
\label{fig:example_cluster_0}
\end{figure*}

\addtocounter{figure}{-1}
\begin{figure*}
\includegraphics[width=0.75\textwidth]{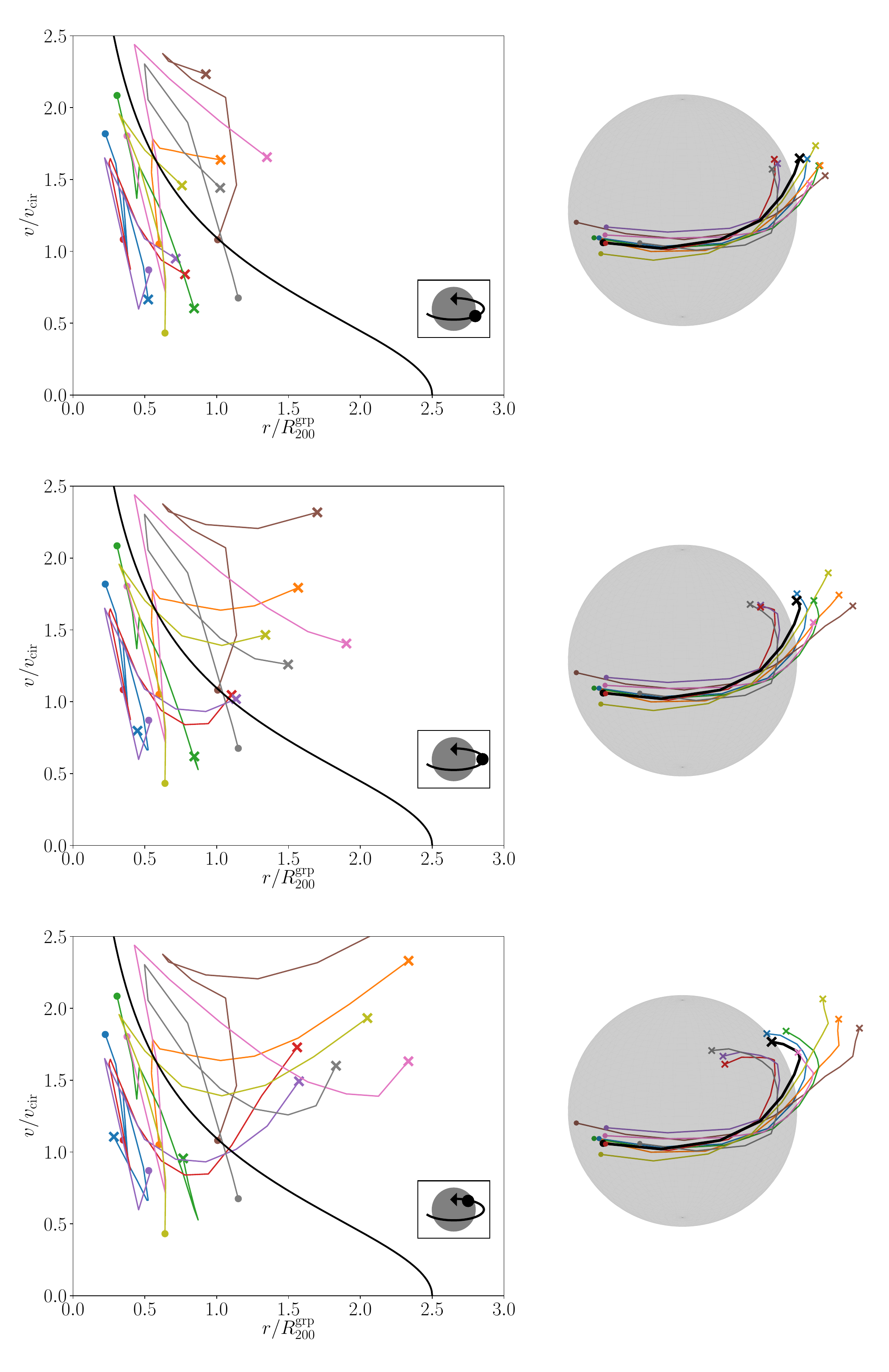}
\caption{{\textit{(continued from previous page)}} Evolution in phase space of an example group.}
\label{fig:example_cluster_1}
\end{figure*}


\bsp	
\label{lastpage}
\end{document}